\documentclass[review,11pt]{elsarticle}

\usepackage{lineno,hyperref}
\usepackage{bm}
\usepackage{algorithm}
\usepackage{algorithmicx}
\usepackage[noend]{algpseudocode}
\let\Algorithm\algorithm
\usepackage{caption}
\usepackage{hyperref}
\usepackage[utf8]{inputenc}
\usepackage{lscape}
\usepackage{tikz}
\usepackage{float}
\usepackage{graphicx}
\usepackage{listings}%
\usepackage{amsmath,amsfonts}
\usepackage{wrapfig}
\usepackage{amssymb}
\usepackage{alltt}%
\usepackage{siunitx}
\usepackage{setspace}
\usepackage{booktabs}
\usepackage{eqparbox}
\usepackage{mathtools}
\usepackage[margin=0.5in]{geometry}

\renewcommand\algorithm[1][]{\Algorithm[#1]\setstretch{1.4}}

\usepackage[labelformat=simple]{subcaption} 

\DeclareCaptionLabelFormat{subcaptionlabel}{\normalfont(\textbf{#2}\normalfont)}
\modulolinenumbers[5]

\journal{arXiv}%{Journal of \LaTeX\ Templates}

\begin{document}

\begin{frontmatter}

\title{Solving the Discretised Boltzmann Transport Equations %discrete ordinate equations 
using Neural Networks: Applications in Neutron Transport}

\author[<address link1>]{Toby R.F. Phillips}
\author[<address link1>]{Claire E. Heaney}
\author[<address link1>]{Boyang Chen}
\author[<address link2>]{Andrew G. Buchan}
\author[<address link1>]{Christopher C. Pain}
\address[<address link1>]{{Applied Modelling and Computation Group, Department of Earth Science and Engineering}, {Imperial College London},
{{London}, {SW7 2AZ} {United Kingdom}}}
\address[<address link2>]{{School of Engineering and Materials Science}, {Queen Mary University of London},
{{London}, {E1 4NS} {United Kingdom}}}
%\street{South Kensington Campus}

\begin{abstract}
In this paper we solve the Boltzmann transport equation using AI libraries. The reason why this is attractive is because it enables one to use the highly optimised software within AI libraries, enabling one to run on different computer architectures and enables one to tap into the vast quantity of community based software that has been developed for AI and ML applications e.g. mixed arithmetic precision or model parallelism. 
Here we take the first steps towards developing this approach for the Boltzmann transport equation and  develop the necessary methods in order to do that effectively. This includes:
1) A space-angle multigrid solution method that can extract the level of parallelism necessary to run efficiently on GPUs or new AI computers.
2) A new Convolutional Finite Element Method (ConvFEM) that greatly simplifies the implementation  of high order finite elements (quadratic to quintic, say). 
3) A new non-linear Petrov-Galerkin method that introduces dissipation anisotropically.

\end{abstract}


\end{frontmatter}

% \linenumbers
\nolinenumbers
\section{Introduction}

%***Lets cite this somewhere %\cite{Smith2016}***

Solving the Boltzmann transport equation, an example being for neutron transport, is a challenging problem requiring the solution to have a large number of degrees of freedom in a high-dimensional phase space. 
This phase space typically covers 3D Cartesian space, 
2D in direction or angle of particle travel, 1D in the energy of the particles and 2D in time. A total of 7 dimensions. In this work, we demonstrate a new AI approach to solving the Boltzmann transport equation by considering a problem in 2D Cartesian space and 2D in direction of particle travel. This results in a 4D space and we use a 4D multigrid method as the basis to solve the equations. We also use  an outer iteration, in a single neural network, for the energy group grouping. 
See \cite{Smith2016} for a review of methods for solving the Boltzmann transport equation numerically for problems in nuclear engineering. 

% \textit{Use of AI libraries to solve PDEs:}
AI Libraries have only just begun to be used to solve partial differential equations (PDEs). Previous work includes large Turbulent flow problems, see \cite{Kochov2021,Wang2022}, and moderate Reynolds number flows with interface tracking, see \cite{Zhao2020}, which, using neural networks, have reported excellent results and computational speeds. Others have also used multigrid methods to solve the matrix equations within fluid flow problems with some trainable weights, see \cite{MARGENBERG2022110983}, and again have reported promising results. There is also the untapped promise of being able to exploit the AI software in new ways, for example, the development of large-scale parallel solution methods including the commonly used in AI pipeline parallelism. 
 
The use of AI to solve PDEs is a promising approach as it can lead to both highly efficient implementations as well as models that can run on a wide range of computer architectures, such as CPUs, GPUs and AI computers. 
One can see that it may be a suitable approach when one sees the similarities between discretisation stencils and filters used in convolutional neural networks. 
In principle, this allows the construction of discretisation and solution methods on structured grids, which is the focus of this work. 
One may also extend the approach to unstructured grids using a graph neural network rather than a convolutional approach. 
The current work builds on the work of \cite{Phillips2022diffusion} which uses AI libraries 
to solve the multi-group diffusion equation for nuclear reactor applications and eigenvalue problems. 
A novelty of that work is the use of a single neural network to solve the entire eigenvalue problem. A key novelty of the current work is its extension to solve the Boltzmann transport equation. 

% % \textit{Multigrid Solver:} 
In addition, one can draw on the analogy between structured grid convolutional auto-encoders, used, for example, to compress images, and multigrid methods. 
It turns out that the recently developed U-net architecture, \cite{Ronneberger2015}, is identical to the architecture needed to implement a sawtooth multigrid cycle, \cite{Tsuruga2018}, in neural networks and is the approach used here.  
The Boltzmann equation is solved here in a 4D space-angle space and we thus place a structured grid across this space and produce a series of coarsened grids (as is usual practice, see \cite{MARGENBERG2022110983,Juncai2019}) by coarsening the grid by a factor of two in each of the 4 dimensions using simple prolongation and restriction operators. To converge the solution the sawtooth multigrid cycle is used here, partly because, it often has superior performance to the classical V-cycle and also because it straightforwardly maps onto the architecture of the U-net neural network,  see \cite{Ronneberger2015,IBTEHAZ202074} for U-net applications and architecture. Others have used recurrent U-net architectures\cite{Alom}, like the recurrent network presented here. A single sawtooth cycle, see \cite{Tsuruga2018}, (also a single forward pass through the U-net) requires the formation of the discrete equation residuals on the finest grid and the restriction of this residual onto the series of coarsened grids. Once the coarsest grid level is reached (often one cell/variable) then a Jacobi relaxation can be applied to the coarsened equations with the residual from the finest grid as a source. The result of this is then prolonged onto the next finest grid and the process is repeated right up to the finest grid level, completing one cycle of the multigrid method. 
This cycle, or pass through the U-net,  is repeated to converge the multigrid solution. More elaborate and better converging multigrid cycles exist such as W- and F- cycles or extensions of the basic sawtooth cycle, see \cite{Briggs2000}. However, when running on a GPU, say, with extreme parallelisation one may notice that performing a Jacobi relaxation of the coarsest grid can be just as expensive as on some of the finer grids because there is very limited parallelisation that can be extracted on the coarsest grids. Thus a compromise is necessary between spending more time on the coarser grids (such as with the W-cycle) and less such as with the V or saw-tooth cycles and should take into account the architecture that one wants to run the model on. 

In addition, in our multigrid method, on the finest grid, we pass the residual to an upwind discretisation method on the same grid and as well as on all coarsened grids. This method is used because it is simple, fast and helps stabilise the multigrid solution method. We partly do this because of the difficulties in solving non-symmetric positive definite matrices using multigrid methods which is the case when solving the Boltzmann transport equation in first-order form. However, despite this, major progress has been made on the use of multigrid methods for these equations, see \cite{DARGAVILLE2020109124,DARGAVILLE2021113379}. This is important as the most commonly used approach sweeps, in a serial fashion, through the grid or mesh updating the radiation field and is thus not well suited to run on the highly parallel GPUs or AI computers.

% \textit{New Convolutional Finite Element Method (ConvFEM):} 
One of the most commonly used discretisation methods for the Boltzmann transport equation solution is the Discontinuous Galerkin (DG) method, see original applications to neutron transport in \cite{Reed1973} and more recently for fluids \cite{Cockburn2017}. This is often applied in linear FEM and lumped mass form to Boltzmann transport as used in the Attila code, see \cite{Gifford2006,Mille2016}.  
However, implementing DG using this approach requires a number of neural network channels - one for each node of the finite element (4 for linear quadrilateral elements) - and thus a more complex neural network is needed. 
This increase in complexity of DG further increases for higher-order elements as even more channels are required.  
The complexity increase associated with DG is also shared with continuous elements. 
For example, for quadratic continuous elements, one can see that the stencil is different for the centre and corner and mid-side nodes for quadrilateral elements in 2D and similarly for 3D elements. Thus we develop a new FEM that we call the Convolutional FEM or ConvFEM, for short, that has the same stencil everywhere and thus can result in much simpler neural network implementations. For the linear elements, the ConvFEM approach is identical to the classical bi-linear FEM approach. Other schemes such as that used in the \cite{Royston2018,Hirsch,LEONARD199117} that can use high order polynomials and upwind biases for stability could also be used. This upwind bias means that one should ideally use non-centred filters to implement them as efficiently as possible. 

% \textit{Non-Linear Petrov-Galerkin Methods:} 
However, the solution of the Bubnov-Galerkin discretisations of the Boltzmann transport equations is infamously unstable, for first-order transport equations, which is also the case for 
similar diamond differencing schemes. 
Our solution to this is to add a non-linear Petrov-Galerkin term~\citep{HUGHES1986,FANG2013540,Donea}, which is simply expressed as a diffusion term that we add to the overall solution method. 
This not only increases the diagonal dominance of the method but also suppresses Gibbs oscillations, see~\citep{Hirsch} for control volume oscillation suppressing schemes, producing more realistic-looking results that are less prone to produce negative angular fluxes (unphysical solutions). 
Other work on the use of Petrov-Galerkin dissipation for the Boltzmann transport equations includes \cite{Pain2006SpacetimeSU,Merton2009ANO}. 
Here, we develop a slight departure from the original, see \cite{HUGHES1986}, and the above non-linear Petrov-Galerkin methods. This results in anisotropic dissipation.

% \textit{Structure of paper:} 
The sections of the paper are organised as follows.Section~\ref{sec:methodlogy} outlines the methodology for solving and discretising the Boltzmann transport equations, forming the multigrid networks and the filter formation. This is followed by results in section~\ref{sec:results} and finally, section~\ref{sec:conc} contains the conclusions and future work. 

\section{Methodology}\label{sec:methodlogy}
The first part of this section outlines the governing equations and their discretisation. The neural network solver is then explained, along with the 4D multigrid method and how it can be implemented as a neural network. %The power method is then described, which is used to solve eigenvalue problems. 
\subsection{Boltzmann Transport  Equation}\label{diffusion_equation} 
The simplified multi-group steady-state Boltzmann equation for neutron transport can be written as:
\begin{equation}\label{eq:diff-eig}
\begin{split} 
 \Omega \cdot \nabla \psi_g + \Sigma^a_g \psi_g + \sum_{\substack{g^{'}= 1\\ g^{'}\neq g}}^{N_g}\Sigma^s_{g\rightarrow g^{'}}\phi_g &= \lambda\,\chi_g\sum_{g^{'}=1}^{N_g} \nu_{g^{'}} \Sigma^f_{g^{'}} \phi_{g^{'}}+\sum_{\substack{g^{'}= 1\\ g^{'}\neq g}}^{N_g}\Sigma^s_{g^{'}\rightarrow g}\phi_{g^{'}},
\\
& \forall g\in\{1,2,\ldots,N_g\},
 \end{split}
\end{equation}
where $\psi_g$ and $\phi_g$ are the angular and  scalar fluxes of the neutron population, 
$\Omega=(\mu, \nu)^T$ is the direction of particle travel, $\Sigma^a_g$ represents the absorption cross-section, $\Sigma^f_g$ represents the fission cross-section, $\nu_g$ is the average number of neutrons produced per fission event, $\Sigma^s_g$ represents the scatter cross-section, $\chi_g$ is the proportion of neutrons produced for each energy group per fission event, $N_g$ is the number of energy groups used and the subscript $g$ denotes the energy group. The diffusion coefficient, $\Omega=(\mu, \nu)^T$, is the direction of neutron transport. 
The eigenvalue, $\lambda$, is defined as the reciprocal of $k_{\text{eff}}$, that is $\lambda = \frac{1}{k_{\text{eff}}}$, where:
\begin{equation}
 k_{\text{eff}}=\frac{\text{number of neutrons in one generation}}{\text{number of neutrons in the preceding generation}}.
\end{equation}
The boundary condition for reflection is:
\begin{equation}\label{eq:diff-reflect}
\frac{\partial \psi_g}{\partial n} = 0\,,
\end{equation} 
% \todo{should this be $D_g$ and $\phi_g$?}
and for a vacuum or bare surface boundary conditions for incoming, to the domain, directions $(\mu_n, \nu_n)^T$:
\begin{equation}\label{eq:diff-bare} \psi_g(\mu_n, \nu_n) = \psi_{n,g} = 0.
\end{equation} 
% \todo{should this be $D_g$ and $\phi_g$?}
%in which $n$ is the outward-pointing normal to the~boundary.

\subsection{Discrete Ordinates}
\label{DO} 
\begin{figure}[H]
\centering
% top diagrams start ....
\begin{minipage}{.45\textwidth}
  \centering
  \includegraphics[scale=1]{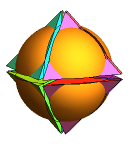}
  \subcaption{Octahedron in a sphere, closed}
  \label{fig:fa_keff_no_cr}
\end{minipage}%
\hfill
\begin{minipage}{.45\textwidth}
  \centering
  \includegraphics[scale=0.5]{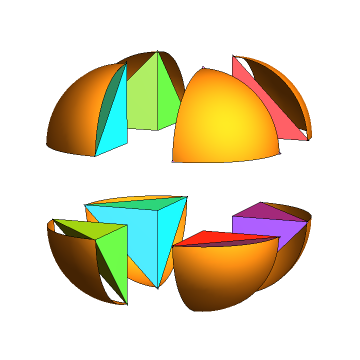}
  \subcaption{Octahedron in a sphere, open}
\end{minipage}
% top diagrams start -----
\begin{minipage}{.33\textwidth}
  \centering
  \includegraphics[scale=0.65]{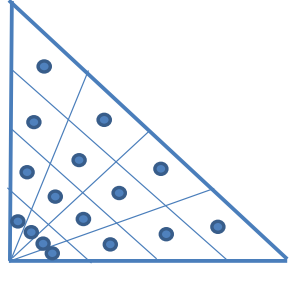}
  \subcaption{$4 \times 4$ grid on a triangle face.}
  \label{fig:triangle_face}
\end{minipage}%
%\hfill
\begin{minipage}{.33\textwidth}
  \centering
  \includegraphics[scale=0.85]{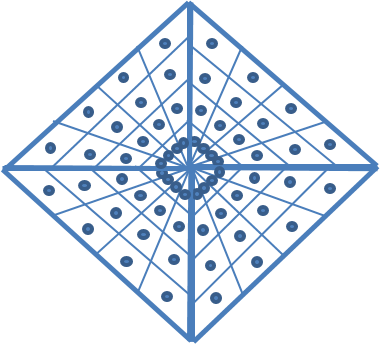}
  \subcaption{Top view of 4 octants placed over half unit sphere. }
  \label{fig:triangle_face_top}
\end{minipage}%
%\hfill
\begin{minipage}{.33\textwidth}
%\hspace{-2.0cm}
  \centering
  \includegraphics[scale=0.55]{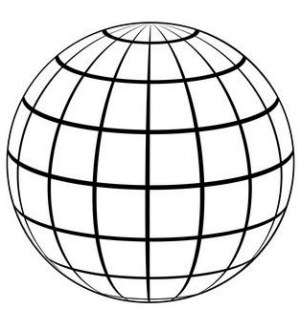}
  \subcaption{Final grid across the unit sphere.}
  \label{fig:unitsphere}
\end{minipage}
\caption{Diagrams showing how the unit sphere is split into 8 parts - forming an octahedron - and then how a regular $N_a\times N_a$ grid is placed over each of those parts or faces. Top shows the octahedron in a sphere, the left hand showing closed and right hand showing open~\citep{octasphere} Bottom shows how a regular grid is placed over the octahedron.
Diagrams showing the grid for a single face and the corresponding unit sphere.}
\label{fig:octa}
\end{figure}

 Here, the set of discrete ordinates that can be used to discretise in angle of particle travel are defined. 
The angle is discretised by placing an octahedron on a unit sphere as can be seen in Figure~\ref{fig:octa}. A rectangular grid is then placed across each face of the octahedron, with a collapsed edge/node, seen in Figure~\ref{fig:triangle_face}. Each octant is then placed together to form the final grid across the unit sphere, shown in Figure~\ref{fig:unitsphere}.

Although a $4 \times 4$ grid is shown, if the grid is kept square (i.e.~$N_a \times N_a$), and $N_a$ is a power of 2, then this is a convenient way for the multigrid approach to be utilised through angle as well as space. Each cell in each face represents a discrete ordinate or SN direction.  
If $N_f=8$ faces are used this results in $N_a \times N_a \times N_f$ discrete ordinate or SN directions. 

Although any number of faces could be used, 8 faces is an ideal number as the directions are similar for every cell within each face as each face occupies an octant. If $\mu_n$, $\nu_n$ and $\xi_n$ represent the angles for a direction $n$ from the x-axis, y-axis and z-axis respectively of the unit sphere. Each face, therefore, contains directions where all $N_a \times N_a$ directions have the same sign such that:
\begin{eqnarray}
     \mu_n > 0\text{ } \forall\text{ } n&\text{ or }&\mu_n <0\text{ } \forall\text{ } n, \\
     \nu_n > 0\text{ } \forall\text{ } n&\text{ or }&\nu_n <0\text{ } \forall\text{ } n, \\
     \xi_n > 0\text{ } \forall\text{ } n&\text{ or }&\xi_n <0\text{ } \forall\text{ } n,
\end{eqnarray}
where each face contains a different combination of the three angles. Each direction (which is the mean direction over the patches shown after projecting onto the unit sphere) also has a corresponding weight or area of the patch associated with it ($p_n$). 
On the unit sphere, assuming coordinates $(\mu, \nu, \xi)$,  the mean directions are given, for patch $n$, by:
\begin{equation}
   \mu_n = \frac{\int_{S_n} \mu dS}{\int_{S_n} dS},
   \;\;\;\;\;  
   \nu_n = \frac{\int_{S_n} \nu dS}{\int_{S_n} dS},
   \;\;\;\;\;  
   \xi_n = \frac{\int_{S_n} \xi dS}{\int_{S_n} dS}, 
\end{equation}
and $p_n=\int_{S_n} dS$ and 
in which $S_n$ is the $n^\text{th}$ surface patch on the unity sphere, see Figure~\ref{fig:unitsphere}. 
For a 2D system $\xi_n = 0$ for all $N$. 
Some of the inaccuracies in the surface area calculation (because of the quadrature used) are accounted 
for by normalising the weights $p_n$ such that:

\begin{equation}
    \sum^N_{n=1}p_n= 4\pi. 
\end{equation}
%\subsection{Difference equation}
After integrating over each path of the unit sphere $n$ and then dividing through by the area of the patch $p_n$, for a given direction $n$ and energy group $g$, Equation~\ref{eq:diff-eig} becomes :
\begin{equation}\label{gov-simple}%\label{eq:disc_ang_hfm}
    \left[ \mu_n\frac{\partial}{\partial x} + \nu_n\frac{\partial}{\partial y} + \tilde\sigma_{Tn,g} \right ]\tilde\psi_{n,g} = \tilde{q}_{n,g}, \;\;\; \forall n\in\{1,2,...,N\}, \;\;\; \forall g\in\{1,2,...,N_g\},
\end{equation}
in which $\mu_n$ is the angle relative to the x-axis, $\nu_n$ is the angle relative to the y-axis, $\tilde\sigma_{Tn,g} $ is the total cross section for direction $n$ and energy group $g$, $\tilde\psi_{n, g}$ is the angular flux for direction $n$ and energy group $g$ and $\tilde{q}_{n, g}$ is the source  of neutrons. 
The discrete ordinate directions $(\mu_n, \nu_n)^T$ is formed from the average direction associated with each patch $n$ on the unit sphere. The angular flux ($\psi$)  and emission of neutrons ($q$) are determined from the scalar flux ($\phi$) and representative source term ($s$,  Equation~\ref{eq:diff-eig}) respectively:

\begin{equation}
    \tilde\psi_{n,g} = p_{n} \tilde\phi_{g}/p_n  = \tilde\phi_{g}\text{ , }\;\;\; \tilde{q}_{n,g} = p_{n} s_{g}/p_n = s_g, 
\end{equation}
and we have divided through by $p_n$ because we have divided the entire equation
by $p_n$ and for an isotropic source for group $g$ of $s_g$.

For a multi-group problem, assuming isotropic scattering, the 
total cross-section of neutrons $\tilde\sigma_{Tn,g} $  at a node 
represented by indexes $i,j$, and for direction $n$ and energy group $g$, that is  $\sigma_{Ti,j,n,g}=(\tilde\sigma_{Tn,g})_{i,j} $, is defined using: 
\begin{equation}
    \sigma_{Ti,j,n,g} =  {\Sigma^a}_{i,j,g} \ + \ \sum_{\substack{g^{'}= 1\\g^{'}\neq g}}^{N_g}\Sigma^s_{i,j,g\rightarrow i,j,g^{'}}, \;\;\;  \forall n\in\{1,2,...,N_a\}. 
\end{equation}
After the angular flux in all $N_d$ directions has been determined, the scalar flux can be calculated using:
\begin{equation}
    \phi_g = \sum^{N_d}_{n=1}p_n\psi_{n,g},
\end{equation}
in which the quadrature weights are equal to the area $p_n$ of each patch $n$ on the unit sphere associated with direction $n$, see Figure \ref{fig:octa}. 
The code for the quadrature sets including the directions and 
weights (patch areas) can be found in the GitHub repository 
\cite{Phillips2022NNTS}.

\subsection{Discretisation in Cartesian space}
An upwind control-volume discretisation of the Boltzmann transport  equation in 2D with 
a regular mesh of $N_x \times N_y$ nodes can be written as:
% cheating a bit with eqnarray
\begin{eqnarray}\label{eq:discretised_hfm}
& & {\cal H}(\mu_n) \mu_d\frac{({\psi}_{i,j,n,g}-{\psi}_{i-1,j,n,g})}{\Delta x}  + 
   {\cal H}(-\mu_n) \mu_n\frac{({\psi}_{i-1,j,n,g}-{\psi_d}_{i,j,n,g})}{\Delta x} 
   \nonumber\\[2mm] 
 & & +%\label{eq:discretised_hfm} 
{\cal H}(\nu_n) \nu_n\frac{({\psi_d}_{i,j,g}-{\psi_d}_{i,j-1,g})}{\Delta y}  +
   {\cal H}(-\nu_n) \nu_n\frac{({\psi}_{i,j-1,n,g}-{\psi}_{i,j,n,g})}{\Delta y}  
   \nonumber\\[2mm] 
 & & +%\label{eq:discretised_hfm}
 {\Sigma^a}_{i,j,n,g}\phi_{i,j,n,g} \ + \ \sum_{\substack{g^{'}= 1\\g^{'}\neq g}}^{N_g}\Sigma^s_{i,j,n,g\rightarrow i,j,g^{'}}\phi_{i,j,n,g}
= \sum_{g^{'}=1}^{N_g}\Sigma^s_{i,j,g^{'}\rightarrow i,j,n,g}\phi_{i,j,g^{'}}
+
\lambda\chi_g\sum_{g^{'}=1}^{N_g} { \nu_{fis}}_{g^{'}} \Sigma^f_{i,j,g^{'}} \phi_{i,j,n,g^{'}}, \label{upwind-eqn} \\[2mm]
& & \qquad \forall i\in\{l,3,..,N_x-l\},\qquad \forall j\in\{l,3,..,N_y-l\},\qquad \forall g\in\{1,2,..,N_g\}, 
\qquad \forall n\in\{1,2,..,N_d\}, 
\nonumber
\end{eqnarray}
in which subscript $n$ represents direction, ${\cal H}(\nu_n)=1$ if $\nu_n>0$ else ${\cal H}(\nu_n)=0$, 
$\Delta x$ and $\Delta y$ are the uniform cell or node widths in the $x$ and $y$ directions respectively, $N_x$ and $N_y$ are the numbers of nodes in the $x$ and $y$ directions respectively, the subscripts~$i$ and $j$ refer to the cells in the $x$ and $y$ directions respectively, $N_g$ is the number of energy groups, the subscript $g$ refers to the energy group and $\phi_{i,j,n,g}$ represents the scalar flux in energy group $g$ in cell or node~$i,j$. 

For the incoming zero angular flux directions, we set the halo values, around the domain, to zero (bare surface boundary condition, see Equation~\eqref{eq:diff-bare}). 
For outgoing flux (no boundary condition needed) we fill the halo nodes with the values of the solution just inside the domain and next to the boundary. 
This approach enables us to have the same stencil everywhere and results in a highly efficient implementation. For the upwind scheme, this adjustment to the halo values does not change the solution (because of the full upwind bias of the discretisation) but it does affect the solution from the Petrov-Galerkin method (below). 

The Petrov-Galerkin method, see \cite{donea2003finite}, applied to Equation~\eqref{gov-simple} can be expressed, in filter form, as:
\begin{equation}
 \bm{r}_{n,g} =
\bm{f}(\bm{\Psi}_{n,g};\mu_n\bm{w_{x}})
+
\bm{f}(\bm{\Psi}_{n,g};\nu_n \bm{w_{y}})
+ 
 \bm{f}^{\text{k-Diff}}(\bm{\Psi}_{n,g};\bm{{k_x}}_{n,g},\bm{w_{Diffxx}})
+ 
 \bm{f}^{\text{k-Diff}}(\bm{\Psi}_{n,g};\bm{{k_y}}_{n,g},\bm{w_{Diffyy}})
 -
 \bm{f}(\bm{q}_{n,g};\bm{w_{ml}}) = \bm{0},
 \label{Petrov}
\end{equation}
in which 
$\bm{\Psi}_{n,g}$ is a $N_x \times N_y$ matrix containing all $\psi_{i,j,n,g}$ components for angle $n$ and group $g$, $\bm{f}$ represents the filter operations for advection (first two terms of equation \eqref{gov-simple}) and is a matrix of size $N_x\times N_y$, for example at node $i,j$ its 
component values are:
\begin{align}
%\begin{equation}
\bm{f}(\bm{\Psi}_{n,g};\mu_n\bm{w_{x}})\vert_{i,j}
= \sum_{u=-l}^l\sum_{v=-l}^l \mu_n {w_{x}}_{u,v} {\Psi}_{i,j,n,g} 
,  \label{conv-wx}\\
%\;\;\;
\forall n\in\{1,2,...,N_d\}, \; 
\forall g\in\{1,2,...,N_g\}, \; 
\forall i\in\{1+l,2,...,N_x-l\}, \; 
\forall j\in\{1+l,2,...,N_y-l\}, \; 
%\end{equation}
\end{align}
and $\bm{f}^{\text{k-Diff}}$, in equation \eqref{Petrov}, are  the diffusion filter operations for the numerically stabilising diffusion used in the Petrov-Galerkin method - noticed that this is applied anisotropically in the x- and y-directions from the non-linear stabilisation within the Petrov-Galerkin method (see next section). 
$ \bm{q}_{n,g}$ in Equation~\eqref{Petrov} contains the scatter and removal operator as well as the fission source from the eigenvalue problem r.h.s. of Equation~\eqref{disc_fiss}. We often collect the 
scalars $\mu_n$, $\nu_n$, $\forall n$ into 
the vectors $\bm{\mu}=(\mu_1 \; \mu_2 \; ...\;\mu_{Nd})^T$, 
$\bm{\nu}=(\nu_1\; \nu_2\; ...\;\nu_{Nd})^T$. 
We use a lumped approximation for this term and thus $\bm{w_{ml}}$ is a $1\times1$ filter containing the mass $\Delta x\Delta y$ associated with all the nodes. The other filters 
$\bm{w_x}$, $\bm{w_y}$, $\bm{w_{Diffxx}}$, $\bm{w_{Diffyy}}$ are matrices of size $(2l+1)\times(2l+1)$ and are the result of discretising the operators $\frac{\partial}{\partial x}$, $\frac{\partial}{\partial y}$, $\frac{\partial^2}{\partial x^2}$, $\frac{\partial^2}{\partial y^2}$ respectively, see Appendix for definitions for various ConvFEM discretisations of these operators. 
For computational speed, it can be more efficient to use 
$\bm{f}(\bm{\Psi}_{n,g};\mu_n\bm{w_{x}} +\nu_n \bm{w_{y}} )$ 
rather than
$\bm{f}(\bm{\Psi}_{n,g};\mu_n\bm{w_{x}})
+
\bm{f}(\bm{\Psi}_{n,g};\nu_n \bm{w_{y}})$ in equation \eqref{Petrov}.  
In the above  $\bm{r}_{n,g}$  is the discrete equation residual and the source $\bm{q}_{n,g}$ effectively gathers all the 
terms without spatial gradients in Equation~\eqref{upwind-eqn}.  

The diffusion calculation is formed for directions x- and y-  using the identities:
\begin{align}%\label{eq:conv_filter_version}
  - {\frac{\partial}{\partial x}} \left({k_x}_{n,g} 
  \frac{\partial \tilde\psi_{n,g}}{\partial x}\right) & = -\frac{1}{2}\left( \frac{\partial^2}{\partial x^2}({k_x}_{n,g} \tilde\psi_{n,g}) + {k_x}_{n,g}  \frac{\partial^2}{\partial x^2}\tilde\psi_{n,g} - \tilde\psi_{n,g} \frac{\partial^2}{\partial x^2}
  {k_x}_{n,g}\right) && \textit{analytical form}\label{eq:analytical_form_x}\\
  &\sim  \bm{f}(\bm{{k_x}}_{n,g}\odot\bm{\Psi}_{n,g};\bm{w_{Diffxx}}) + \bm{{k_x}}_{n,g}\odot \bm{f}(\bm{\Psi}_{n,g};\bm{w_{Diffxx}}) &&\nonumber \\
  &\quad - \bm{\Psi}_{n,g}\odot \bm{f}(\bm{{k_x}}_{n,g};\bm{w_{Diffxx}}) &&\nonumber \\
  &\quad =  \bm{f}^{\text{k-Diff}}(\bm{\Psi}_{n,g};\bm{{k_x}}_{n,g},\bm{w_{Diffxx}}) \; , && \textit{discretised form}
  \label{kxx}
\end{align}
and 
\begin{align}%\label{eq:conv_filter_version}
  - {\frac{\partial}{\partial y}} \left({k_y}_{n,g} 
  \frac{\partial \tilde\psi_{n,g}}{\partial y}\right) & = -\frac{1}{2}\left( \frac{\partial^2}{\partial x^2}({k_y}_{n,g} \tilde\psi_{n,g}) + {k_y}_{n,g}  \frac{\partial^2}{\partial y^2}\tilde\psi_{n,g} - \tilde\psi_{n,g} \frac{\partial^2}{\partial y^2}
  {k_y}_{n,g}\right) && \textit{analytical form}\label{eq:analytical_form_y}
  \\
  &\sim  \bm{f}(\bm{{k_y}}_{n,g}\odot\bm{\Psi}_{n,g};\bm{w_{Diffyy}}) + \bm{{k_y}}_{n,g}\odot \bm{f}(\bm{\Psi}_{n,g};\bm{w_{Diffyy}}) &&\nonumber \\
  &\quad - \bm{\Psi}_{n,g}\odot \bm{f}(\bm{{k_y}}_{n,g};\bm{w_{Diffyy}}) &&\nonumber \\
  &\quad = \bm{f}^{\text{k-Diff}}(\bm{\Psi}_{n,g};\bm{{k_y}}_{n,g},\bm{w_{Diffyy}}) \; ,&& \textit{discretised form}
  \label{kyy}
\end{align}
where $\odot$ denotes the Hadamard product which performs entrywise multiplication, $\bm{{k_x}}_{n,g}$, $\bm{{k_y}}_{n,g}$ are $N_x \times N_y$ matrices containing all diffusion values in the x- and y-directions ${k_x}_{i,j,n,g}$, ${k_y}_{i,j,n,g}$ and $\bm{f}^{\text{k-Diff}}$ represents the application of the convolutional layer with weights associated with the discretised Laplacians in the x-direction $\frac{\partial^2}{\partial x^2} \sim  \bm{w_{Diffxx}}$ and y-direction $\frac{\partial^2}{\partial y^2} \sim  \bm{w_{Diffyy}}$. The equations \eqref{kxx},\eqref{kyy} act as a definition of $\bm{f}^{\text{k-Diff}}$ 
and show how it is used to form second order derivatives.

% For bare surface boundary conditions (Equation~\eqref{eq:diff-bare}), where the normal to the boundary is aligned with the $x$-direction, the absorption term is modified by:
% \begin{equation}\label{eq:abs-bc-x}
% \Sigma^a_{i,j,g}\leftarrow \Sigma^a_{i,j,g} + \frac{1}{2\Delta x}\,.
% \end{equation}
% To apply this to where the normal to the boundary is aligned with the $y$ direction: 
% \begin{equation}\label{eq:abs-bc-y}
% \Sigma^a_{i,j,g}\leftarrow \Sigma^a_{i,j,g} + \frac{1}{2\Delta y}\,.
% \end{equation}
% For cells that have both boundary conditions the following modification is made: 
% \begin{equation}\label{eq:abs-bc-xy}
% \Sigma^a_{i,j,g}\leftarrow \Sigma^a_{i,j,g} + \frac{1}{2\Delta x} + \frac{1}{2\Delta y}\,. 
% \end{equation}
The discretised form of Equation~\eqref{eq:diff-eig} can therefore be written, in matrix from, as:
\begin{equation}
 \bm{A}\bm{\phi}=\lambda \bm{B}\bm{\phi}.
 \label{disc_fiss}
\end{equation}
in which $\bm{\phi}$ is the entire 
solution vector and thus the 
matrix~$\bm{A}$ contains the absorption, diffusion and scattering out of energy groups from the left-hand side of Equation~\eqref{eq:discretised_hfm}, matrix~$\bm{B}$ represents the fission terms in the right-hand side of Equation~\eqref{eq:discretised_hfm} and the vector~$\bm{\phi}$ contains the values of the scalar flux for each cell in every energy group. The~matrices are of size $N_g(N_x-l)(N_y-l)$ by $N_g(N_x-l)(N_y-l)$ in which we have not counted the halo nodes. Although~a 2D discretisation is given here, the~methods could be applied in 1D or~3D.

\subsection{Neural network filters }
\label{sec:nn-alt}

\begin{figure}[!htb]
\centering
\scalebox{0.5}{\begin{tikzpicture}
\draw[step=1.0,black,thin] (0,0) grid (5,5);
\draw[blue,line width = 1mm] (1,1) rectangle ++ (3,3);
\draw[red,line width = 1mm] (2,2) rectangle ++ (1,1);
\foreach \x in {0.5,...,4.5}
        \node[font=\huge] at (0.5,\x) {1}; 
\foreach \x in {0.5,...,4.5}
        \node[font=\huge] at (1.5,\x) {2}; 
\foreach \x in {0.5,...,4.5}
        \node[font=\huge] at (2.5,\x) {5}; 
\foreach \x in {0.5,...,4.5}
        \node[font=\huge] at (3.5,\x) {4}; 
\foreach \x in {0.5,...,4.5}
        \node[font=\huge] at (4.5,\x) {1}; 

\draw[step=1.0,black,thin] (7,1) grid (10,4);
\node[] at (2.5,6) {Input (Nx x Ny)};
\node[] at (8.5,6) {Filter};
\node[font=\huge] at (7.5,1.5) {0}; 
\node[font=\huge] at (7.5,2.5) {-1}; 
\node[font=\huge] at (7.5,3.5) {0}; 
\node[font=\huge] at (8.5,1.5) {-1}; 
\node[font=\huge] at (8.5,2.5) {4}; 
\node[font=\huge] at (8.5,3.5) {-1}; 
\node[font=\huge] at (9.5,1.5) {0}; 
\node[font=\huge] at (9.5,2.5) {-1}; 
\node[font=\huge] at (9.5,3.5) {0};

\node[font=\huge] at (6,2.5) {*};
\node[font=\huge] at (11,2.5) {=};
\draw[step=1.0,black,thin] (12,1) grid (15,4);
\node[] at (13.5,6) {Sum of Values};
\node[font=\huge] at (12.5,1.5) {0}; 
\node[font=\huge] at (12.5,2.5) {-2}; 
\node[font=\huge] at (12.5,3.5) {0}; 
\node[font=\huge] at (13.5,1.5) {-5}; 
\node[font=\huge] at (13.5,2.5) {20}; 
\node[font=\huge] at (13.5,3.5) {-5}; 
\node[font=\huge] at (14.5,1.5) {0}; 
\node[font=\huge] at (14.5,2.5) {-4}; 
\node[font=\huge] at (14.5,3.5) {0}; 
\draw[blue,line width = 1mm] (12,1) rectangle ++ (3,3);
\node[font=\huge] at (16,2.5) {=};
\node[] at (17.5,6) {Updated Red Value};
\draw[red,line width = 1mm] (17,2) rectangle ++ (1,1);
\node[font=\huge] at (17.5,2.5) {4}; 
\end{tikzpicture}}
\caption{A convolutional filter which applies the discretised diffusion operator (a five-point finite-difference stencil) in 2D to 9 cells or grid points. The filter is first applied to all the cells in the blue block on the left; the result of which can be seen in the blue block after the equals sign. The 9 values are then summed to give the value in the red block on the right which is the output value.}
\label{fig:conv_filter}
\end{figure}
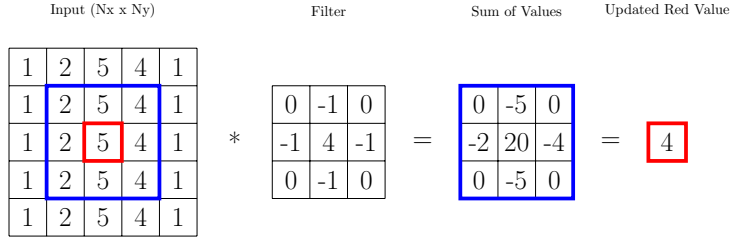

This section describes the neural network filter and shows how they can implemented as a convolutional layer of a neural network using pre-determined weights. All networks were implemented in python using Keras~\cite{chollet2015keras} with the TensorFlow backend~\cite{tensorflow2015}. A convolutional layer has a filter or kernel, which is a small grid (smaller than the input data and typically of dimension $3\times 3$, $5\times 5$ or $7\times 7$) whose cells have values known as weights associated with them. The filter is applied to part of the input by multiplying the input value by the weight in the overlapping cells. The products are summed to produce the output. The filter is then applied to a neighbouring part of the input and another output is created. This process is repeated until the filter has passed over all the input data. The action of a 2D convolutional layer on a 2D input can be written as follows as shown in 
Equation~\eqref{conv-wx}. 
In compact form this is generally: 
\begin{equation}
\bm{a}^{k+1}_{n,g} =  \bm{f}(\bm{a}^{k}_{n,g}; \bm{w} ).
\end{equation}
The weights of the filter are represented by $w_{u,v}$ and are contained in the filter in matrix form $\bm{w}$ and, in this case, the size of the filter is $(2l+1) \times (2l+1)$.  A filter acting on one piece of input data, with $l=1$, can be seen in Figure~\ref{fig:conv_filter}. %and $w_{u,v}$ is the $uv$ component of a $(2l+1) \times (2l+1)$ matrix containing the weights and $x_{i,j}$ is the $ij$ component in a 2D matrix, where superscript $k$ is increased by one when a convolutional pass is done through the entire matrix.

% \todo{we should explain $x$.}
%The advection terms in equation~\eqref{upwind-eqn} can be resolved using convolutional passes with filters:
%\begin{equation}\label{eq:conv_advection_term}
%     \mu_n\frac{\partial \bm{\Psi_n}}{\partial x} + \nu_n\frac{\partial\bm{\Psi_n}}{\partial y} = \mu_n 
%     f^{\text{XAdv}}(\bm{\Psi_n};\bm{w_{x}}) +\nu_nf^{\text{YAdv}}(\bm{\Psi_n};\bm{w_{y}})
%\end{equation}

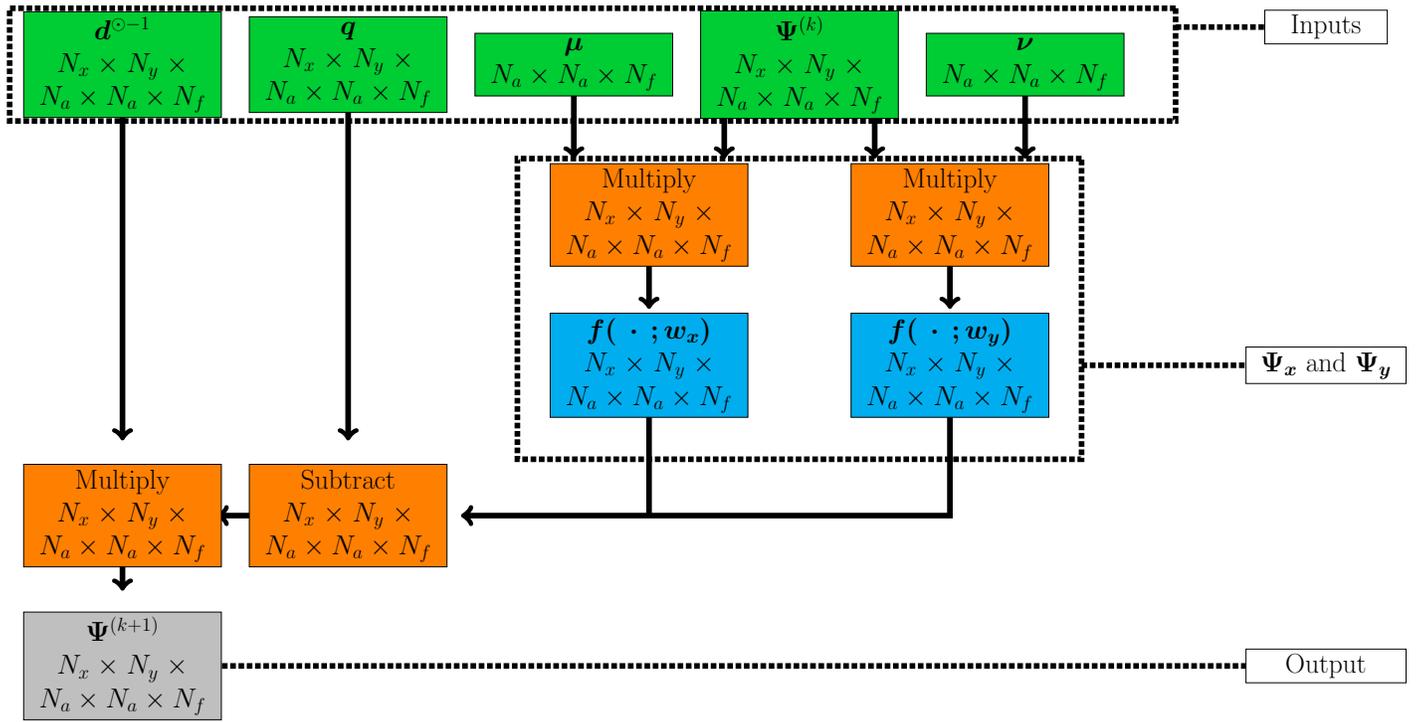
\begin{figure}[!htb]
\centering
\scalebox{0.5}{\begin{tikzpicture}
% \filldraw[blue] (0,6) circle (8pt);
% \filldraw[blue] (0,10) circle (8pt);
% \draw [line width=1.5ms-to] (8,12) -- (8,9);
% \draw [line width=1.5mm,-to] (9,12) -- (9,10) -- (11,10) -- (11,9);

% \draw [line width=1.5mm,-to] (12,8) -- (12,5);
% \draw [line width=1.5mm,-to] (8,8) -- (8,5);

\draw [line width=1.5mm,-to] (0,12) -- (0,2);
\draw [line width=1.5mm,-to] (0,0) -- (0,-2);
\draw [line width=1.5mm,-to] (4,0) -- (2.5,0);
\draw [line width=1.5mm,-to] (6,12) -- (6,2);

\draw [line width=1.5mm,-to] (12,12) -- (12,9.5);
\draw [line width=1.5mm,-to] (24,12) -- (24,9.5);

\draw [line width=1.5mm,-to] (16,12) -- (16,9.5);
\draw [line width=1.5mm,-to] (20,12) -- (20,9.5);

\draw [line width=1.5mm,-to] (22,8) -- (22,5.5);

\draw [line width=1.5mm,-to] (14,8) -- (14,5.5);

\draw [line width=1.5mm,-to] (22,4) -- (22,0) -- (9,0);
\draw [line width=1.5mm] (14,4) -- (14,0);
\draw[line width = 1.5mm, dotted] (28,13.5) -- (28,10.5) -- (-3,10.5) -- (-3,13.5) -- (28,13.5);
\draw[line width = 1.5mm, dotted] (28,13) -- (32,13);
\node[draw,font=\huge,text width = 3cm,align=center, fill = white] at (32,13) {Inputs};
\draw[line width = 1.5mm, dotted] (25.5,9.5) -- (10.5,9.5) -- (10.5,1.5) -- (25.5,1.5) -- (25.5,9.5);
\draw[line width = 1.5mm, dotted] (25.5,4) -- (32,4);
\node[draw,font=\huge,text width = 4cm,align=center, fill = white] at (32,4) {$\bm{\Psi}_{\bm{x}}$ and $\bm{\Psi}_{\bm{y}}$};
\draw[line width = 1.5mm, dotted] (0,-4) -- (32,-4);
\node[draw,font=\huge,text width = 4cm,align=center, fill = white] at (32,-4) {Output};
% \draw [line width=1.5mm,-to] (8,4) -- (8,0.5) -- (5.75,0.5);
% \draw [line width=1.5mm,-to] (12,12) -- (15,12) -- (15,6) -- (9,6) -- (9,5);

% \draw [line width=1.5mm,-to] (0,0) -- (0,-2.7);

\node[draw,font=\huge,text width = 5cm,align=center,fill=green!80!blue] at (18,12) {$\bm{\Psi}^{(k)}$ \\ $N_x\times N_y\times N_a\times N_a\times N_f$};

\node[draw,font=\huge,text width = 5cm,align=center,fill=green!80!blue] at (12,12) {$\bm{\mu}$ \\ $N_a\times N_a\times N_f$};
\node[draw,font=\huge,text width = 5cm,align=center,fill=green!80!blue] at (24,12) {$\bm{\nu}$ \\ $N_a\times N_a\times N_f$};
\node[draw,font=\huge,text width = 5cm,align=center,fill=green!80!blue] at (6,12) {$\bm{q}$ \\ $N_x\times N_y\times N_a\times N_a\times N_f$};
\node[draw,font=\huge,text width = 5cm,align=center,fill=green!80!blue] at (0,12) {$\bm{d}^{\odot-1}$ \\ $N_x\times N_y\times N_a\times N_a\times N_f$};

\node[draw,font=\huge,text width = 5cm,align=center,fill=orange] at (14,8) {Multiply \\$N_x\times N_y\times N_a\times N_a\times N_f$};

\node[draw,font=\huge,text width = 5cm,align=center,fill=cyan] at (14,4) {$\bm{f(\;\cdot\;;\bm{w_x})}$  \\ $N_x\times N_y\times N_a\times N_a\times N_f$};

\node[draw,font=\huge,text width = 5cm,align=center,fill=orange] at (22,8) {Multiply \\ $N_x\times N_y\times N_a\times N_a\times N_f$};

\node[draw,font=\huge,text width = 5cm,align=center,fill=cyan] at (22,4) {$\bm{f(\;\cdot\;;\bm{w_y})}$  \\ $N_x\times N_y\times N_a\times N_a\times N_f$};

\node[draw,font=\huge,text width = 5cm,align=center,fill=orange] at (6,0) {Subtract \\ $N_x\times N_y\times N_a\times N_a\times N_f$};

\node[draw,font=\huge,text width = 5cm,align=center,fill=orange] at (0,0) {Multiply \\ $N_x\times N_y\times N_a\times N_a\times N_f$};

\node[draw,font=\huge,text width = 5cm,align=center,fill=lightgray] at (0,-4) {$\bm{\Psi}^{(k+1)}$  \\ $N_x\times N_y\times N_a\times N_a\times N_f$};

\end{tikzpicture}}

\caption{This is a schematic of the Neural network, $J$, representing a single Jacobi iteration. This network performs a single Jacobi iteration on the flux of a single energy group. Takes the angular flux ($\bm{\Psi}^{(k)}$), source ($\bm{q}$), angular velocities ($\bm{\mu}$,$\bm{\nu}$) and the strictly diagonal coefficients ($\bm{d}_{}^{\odot-1}$) as inputs (green boxes). A number of layer operations are performed, mathematical operations in orange and convolutional passes in cyan. Finally, it outputs the flux of the next Jacobi iterations ($\bm{\Psi}^{(k+1)}$). Arrow origins show which layer the data originated and the end of the arrow shows which layer takes that data as input. Dimensions of layers are given on the second line of each box.}
\label{fig:jacobi_net}
\end{figure}

%where $f^{\text{XAdv}}$ and $f^{\text{YAdv}}$ are convolutional passes representing the advection in the x and y-direction respectively. These passes have weights matrices ($\bm{w_x}$ and $\bm{w_y}$) associated with them. 
For the upwind approximation of advection (see Equation~\eqref{eq:discretised_hfm}), the filter weights have different values, based on the the signs of $\mu_n$ and $\nu_n$. For the x-direction:
\begin{gather}
\bm{w_{x}}
 = 
 \begin{bmatrix}
 0
 & 0 & 0 \\
\frac{-1}{\Delta x} & \frac{1}{\Delta x} &0 \\
 0 & 0 & 0
 \end{bmatrix} \text{ if }\mu_n > 0,
\end{gather}
or
\begin{gather}
\bm{w_{x}}
 = 
 \begin{bmatrix}
 0
 & 0 & 0 \\
0&\frac{-1}{\Delta x} & \frac{1}{\Delta x} \\
 0 & 0 & 0
 \end{bmatrix} \text{ if }\mu_n < 0,
\end{gather}
and for the y-direction:

\begin{gather}
\bm{w_{y}}
 = 
 \begin{bmatrix}
 0
 & 0 & 0 \\
0 & \frac{1}{\Delta y} &0 \\
 0 & \frac{-1}{\Delta y} & 0
 \end{bmatrix} \text{ if }\nu_n > 0,
\end{gather}
or
\begin{gather}
\bm{w_{y}}
 = 
 \begin{bmatrix}
 0
 &  \frac{1}{\Delta y} & 0 \\
0&\frac{-1}{\Delta y} & 0 \\
 0 & 0 & 0
 \end{bmatrix} \text{ if }\nu_n < 0.
\end{gather}
As previously established, Section~\ref{DO}, every direction in each face has the same sign of $\mu_n$ and $\nu_n$. This means that each face requires one set of filters. A 2D filter is sufficient for a 2D problem with isotropic scattering. If  $\bm{\Psi}_{\bm{x},n,g} =  \bm{f}(\bm{\Psi}_{n,g};\bm{w_{x}}) $ and $\bm{\Psi}_{\bm{y},n,g} =  \bm{f}(\bm{\Psi}_{n,g};\bm{w_{y}})$  is the convolutional discretising the advection in the x- and y-directions, then there are $N_f$ of these filters associated with the discrete ordinate faces shown in Figures \ref{fig:octa}.   Each filter is then applied to all $N_a \times N_a$ directions for a single face. Using the Jacobi method the iteration for direction $n$ and energy group $g$ can be represented as:

\begin{equation}
    \bm{\Psi}_{n,g}^{(k+1)} = \bm{d}_{n,g}^{\odot-1} \odot (   \bm{d}_{n,g}\odot \bm{\Psi}^{(k)}_{n,g} -\bm{r}^{(k)}_{n,g}),
    \label{eq:matrix_full}
\end{equation}
in which the vector $\bm{d}_{n,g}$ contains the diagonal of the 
matrix or filter upwind discretised system multiplied (see Equation~\eqref{eq:discretised_hfm}) by a factor $\beta$ ($\beta=3$ is used here) to introduce relaxation into the scheme on the finest grid when the Petrov-Galerkin discretisation is applied. The inverse Hadamard product~\citep{Reams1999}, used in the above, is defined 
$\bm{d}_{n,g}^{\odot-1} \vert_{i,j} = \frac{1}{ d_{i,j,n,g}}$. The superscript $k$ refers to the iteration level and the residual $\bm{r}^{(k)}_{n,g}$ is formed from 
equation \eqref{Petrov} with the most recent value of the angular flux $\bm{\Psi}^{(k)}_{n,g}$. 
When the upwind scheme is applied on the finest grid and on all coarsened grid levels $\beta=1$ is used. 
Figure~\ref{fig:jacobi_net} details the architecture of the neural network that performs the Jacobi iteration given by 
Equation~\eqref{eq:matrix_full}.

%\clearpage

\subsection{Space-Angle Multigrid}\label{sec:multi-grid}

\begin{figure}[!htb]
\centering
\scalebox{0.7}{\includegraphics{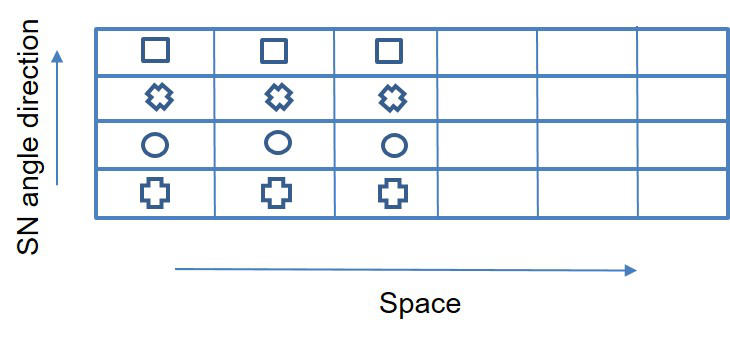}}
\caption{Space and Angle multigrid schematic showing how the multigrid method works. 
The different shapes show that the filters are different for each angle direction or ordinate (SN).}
\label{fig:space-angle-multi_grid}
\end{figure}

\begin{figure}[!htb]
\centering
\scalebox{0.5}{\includegraphics{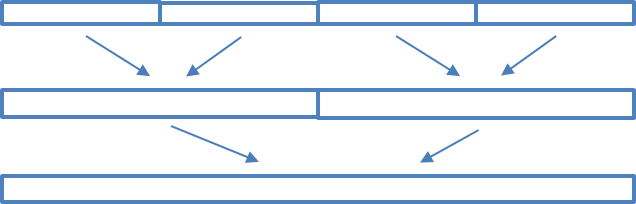}}
\caption{Restriction and prolongation (reverse arrow operation) operations in 1D. In 4D instead of restricting two variables/nodes or cells, one restricts $2^4$ variables in the 4D space and angle multigrid method.}
\label{fig:one-d-restrictions-prolongation.png}
\end{figure}

\begin{figure}[!htb]
\centering
\scalebox{0.44}{\begin{tikzpicture}
\filldraw[cyan] (0,6) circle (8pt);
\filldraw[blue] (0,10) circle (8pt);
% \node[font=\huge] at (0,11) {$\bm{r^{(k)}_1} = \bm{q} - \left[ \bm{\mu_1}\frac{\partial}{\partial x} + \bm{\nu_1}\frac{\partial}{\partial y} + \bm{\sigma}_{T,\bm{1}} \right ]\bm{\Psi_\bm{1}}$};
\node[font=\huge] at (4,11) {$\bm{r}^{(k)}_{\bm{1}}$ calculated using equation ~\eqref{Petrov}};

% ~\eqref{Petrov}
\draw [line width=0.5mm,-to] (0,10) -- (0,8);
\node[font=\huge] at (0,7) {$\bm{r}^{(k)}_{\bm{1}}$};
\draw [line width=1.5mm,-to] (0.2,5.8) -- (3.6,4.2);
\node[font=\huge] at (6,5) {$\bm{r}^{(k)}_{\bm{2}} = \bm{f}^{\text{}}(\bm{r}^{mod,(k)}_{\bm{1}};\bm{w_{R}})$};
\filldraw[cyan] (4,4) circle (8pt);
\draw [line width=1.5mm,-to] (4.2,3.8) -- (7.6,2.2);
\node[font=\huge] at (10,3) {$\bm{r}^{(k)}_{\bm{3}} = \bm{f}^{\text{}}(\bm{r}^{mod,(k)}_{\bm{2}};\bm{w_{R}})$};
\filldraw[cyan] (8,2) circle (8pt);
\draw [line width=1.5mm,-to] (8.2,2) -- (13.6,2);
% \node[font=\huge] at (9.8,2.5) {$J$};

\filldraw[yellow] (14,2) circle (8pt);
% \draw [-to] (8,2) -- (11.6,2);
\filldraw[teal] (18,4) circle (8pt);
\node[font=\huge] at (16,1) {$\Delta \bm{\Psi}^{(k)}_{\bm{3}} = \text{J}\left(\bm{0},\bm{d}_{\bm{3}}^{\odot-1},\bm{s_3}\right)$};
\draw [line width=1.5mm,-to] (14.2,2.2) -- (17.6,3.8);
\node[font=\huge] at (18,5) {$\widetilde{\Delta {\bm{\Psi}}}^{(k)}_{\bm{2}}=\text{UpSamp}(\Delta \bm{\Psi}^{(k)}_{\bm{3}} )$};
\filldraw[yellow] (24,4) circle (8pt);
\draw [line width=1.5mm,-to] (18.2,4) --
(23.6,4);
% \node[font=\huge] at (17.8,4.5) {$J$};
\filldraw[teal] (28,6) circle (8pt);
\draw [line width=1.5mm,-to] (24.2,4.2) -- (27.6,5.8);
\filldraw[yellow] (34,6) circle (8pt);
\draw [line width=1.5mm,-to] (28.2,6) -- (33.6,6);
% \node[font=\huge] at (25.8,6.5) {$J$};
\draw [line width=0.5mm,-to] (0.2,6) -- (27.6,6);
\draw [line width=0.5mm,-to] (4.2,4) -- (17.6,4);
\node[font=\huge] at (26,3) {$\Delta \bm{\Psi}^{(k)}_{\bm{2}} = \text{J}\left(\widetilde{\Delta \bm{\Psi_2}}^{(k)}\bm{d}_{\bm{2}}^{\odot-1},\bm{s_2}\right)$};
\node[font=\huge] at (28,7) {$\widetilde{\Delta \bm{\Psi_1}}^{(k)} =\text{UpSamp}(\Delta \bm{\Psi}^{(k)}_{\bm{2}} )$};
\node[font=\huge] at (36,5) {$\Delta \bm{\Psi^{(k)}_1} = \text{J}\left(\widetilde{\Delta \bm{\Psi_1}}^{(k)},\bm{d}_{\bm{1}}^{\odot-1},\bm{s_1}\right)$};
\filldraw[blue] (34,10) circle (8pt);
\node[font=\huge] at (34,11) {$\bm{\Psi}^{(k+1)_{\bm{1}}} = \bm{\Psi}^{(k)}_{\bm{1}} + \Delta \bm{\Psi}^{(k)}_{\bm{1}}$};
\draw [line width=0.5mm,-to] (34,6.2) -- (34,9.6);
\draw [line width=0.5mm,-to] (34,10) -- (0.4,10);
\end{tikzpicture}}
\caption{A sawtooth cycle multigrid iteration, with the subscript indicating the resolution and superscript representing the multigrid iteration and $\text{J}(\;\cdot\;)$ representing a Jacobi iteration. The residual is calculated and restricted twice. These residuals are used, along with Jacobi iterations, to determine the change in flux. After the finest level is reached, the flux is updated and the process repeats.}
\label{fig:multi_grid_it}
\end{figure}
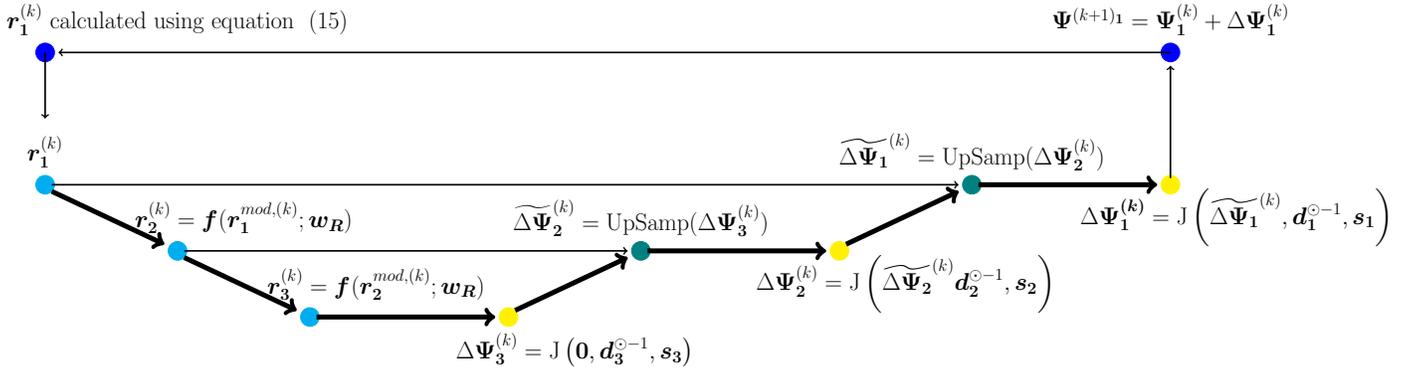

Here we define the 4D multigrid method used here. 4D is generated from 2D in Cartesian space and 2D in angle. The schematic shown in Figure \ref{fig:space-angle-multi_grid} shows how it works and it has the following parts:
(1) We assume in space and in angle (similar in
multi-dimensions) that we have a different filter for each 
discrete ordinate direction but the same across space. 
%2) Add discretization filters together to form coarser
%filters.
(2) In angle we simply add the directions, or patches on the unit sphere, together to form coarser discrete ordinate equations. 
(3) In Cartesian space we discretise on the coarser grids to form the coarse grid equations on each multigrid level. 
It should be noted that the material properties are mapped to a coarser grid using a harmonic average before the discretisation on the coarser grids is formed. 
The same discretisation is used on each multigrid level but with different cell or node sizes $\Delta x$, $\Delta y$. 
The multigrid restriction and prolongation operations associated with this multi-grid method are shown in the schematic~\ref{fig:one-d-restrictions-prolongation.png}. 

We show the multigrid cycle used to form the iterative space-angle multigrid solution method in 
Figure~\ref{fig:multi_grid_it}. This shows a single sawtooth cycle multigrid iteration with two restrictions. 
% The residual ($r_1^k$) is calculated:
% \begin{equation}
%     \bm{r^{(k)}_1} = \bm{q} - \left[ \bm{\mu_1}\frac{\partial}{\partial x} + \bm{\nu_1}\frac{\partial}{\partial y} + \bm{\sigma}_{T,\bm{1}} \right ]\bm{\Psi_\bm{1}}
% \end{equation}
First, the residual from the 
Petrov-Galerkin method, Equation ~\eqref{Petrov} is sent down through all the multigrid levels by restriction. Then the upwind scheme is 
used with the Jacobi iterations on each of the grids starting from the coarsest space-time grid and working up to the finest grid. On the finest grid level, a Jacobi iteration is applied using the upwind scheme and then the Jacobi iteration is applied to the Petrov-Galerkin method but with an enhanced diagonal described by Equation~\eqref{eq:matrix_full}. 
That is, as shown in Figure~\ref{fig:multi_grid_it}, a Jacobi iteration is performed on the lowest level (subscript $3$) to determine $\Delta\bm{\Psi_3}^{(k)}$. This is prolongated to estimate $\widetilde{\Delta \bm{\Psi_2}}^{(k)}$ from which Jacobi smoothing is performed to  obtain ${\Delta \bm{\Psi_2}^{(k)}}$. This repeats until the finest level is reached (subscript $1$), where the flux is updated ($k+1$) and the process is repeated for a number of multigrid iterations.

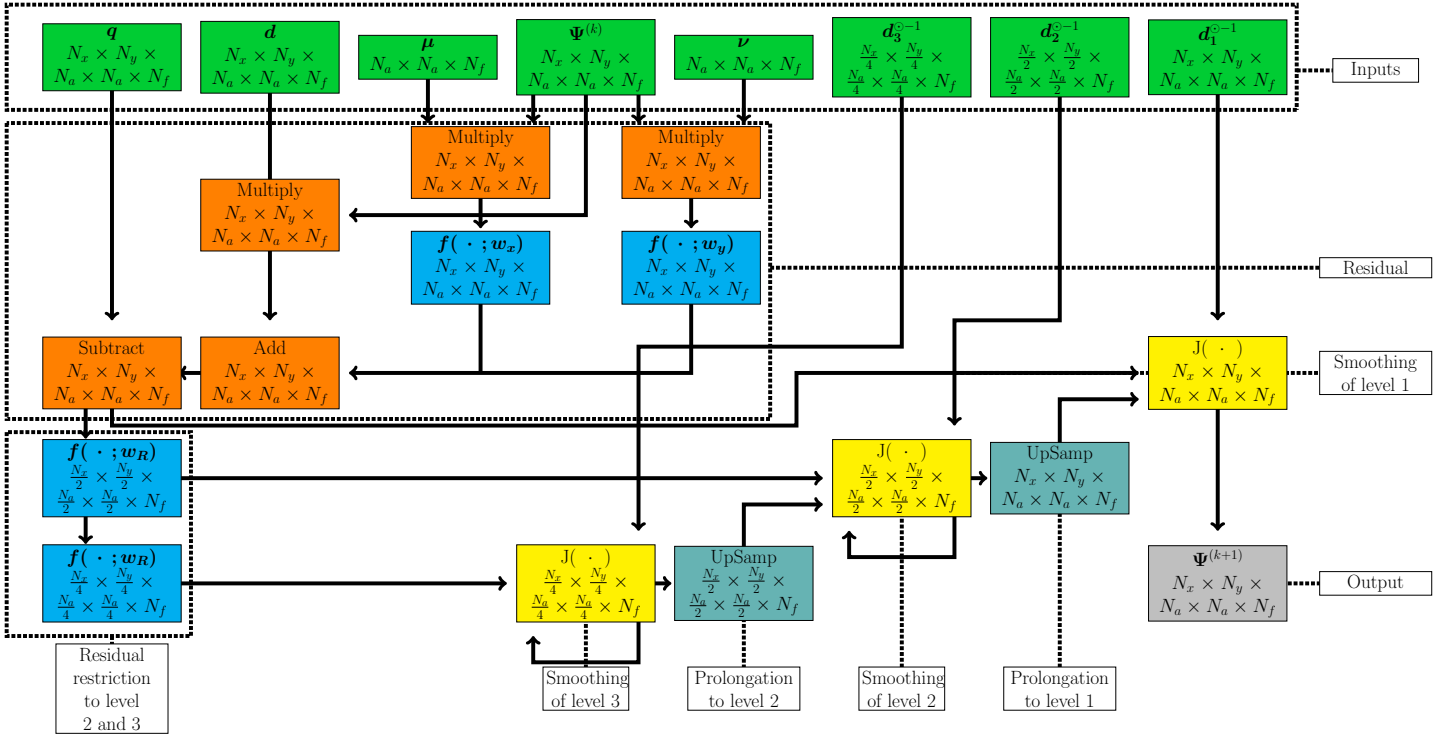
\begin{figure}[!htb]
\centering
\scalebox{0.35}{\begin{tikzpicture}
% \filldraw[blue] (0,6) circle (8pt);
% \filldraw[blue] (0,10) circle (8pt);
% \draw [line width=1.5ms-to] (8,12) -- (8,9);
% \draw [line width=1.5mm,-to] (9,12) -- (9,10) -- (11,10) -- (11,9);

% \draw [line width=1.5mm,-to] (12,8) -- (12,5);
% \draw [line width=1.5mm,-to] (8,8) -- (8,5);

\draw [line width=1.5mm,-to] (0,12) -- (0,2);
\draw [line width=1.5mm,-to] (-1,0) -- (-1,-2.5);
\draw [line width=1.5mm,-to] (4,0) -- (2.5,0);
\draw [line width=1.5mm,-to] (6,12) -- (6,2);

\draw [line width=1.5mm,-to] (12,12) -- (12,9.5);
\draw [line width=1.5mm,-to] (24,12) -- (24,9.5);

\draw [line width=1.5mm,-to] (16,12) -- (16,9.5);
\draw [line width=1.5mm,-to] (20,12) -- (20,9.5);

\draw [line width=1.5mm,-to] (22,8) -- (22,5.5);

\draw [line width=1.5mm,-to] (14,8) -- (14,5.5);

\draw [line width=1.5mm,-to] (22,4) -- (22,0) -- (9,0);
\draw [line width=1.5mm] (14,4) -- (14,0);

\draw [line width=1.5mm,-to] (18,12) -- (18,6) -- (9,6);

\draw [line width=1.5mm,-to] (0,0) -- (0,-2) -- (26,-2) -- (26,0)--(39,0);
\draw [line width=1.5mm,-to] (-1,-4) -- (-1,-6.5);

\draw [line width=1.5mm,-to] (0,-4) -- (27,-4);
\draw [line width=1.5mm,-to] (0,-8) -- (15,-8);

\draw [line width=1.5mm,-to] (20,-8) -- (20,-11) -- (16,-11) -- (16,-10);
\draw [line width=1.5mm,-to] (32,-4) -- (32,-7) -- (28,-7) -- (28,-6);

\draw [line width=1.5mm,-to] (24,-8) -- (24,-5) -- (27,-5);
\draw [line width=1.5mm,-to] (36,-4) -- (36,-1) -- (39,-1);

\draw [line width=1.5mm,-to] (42,12) -- (42,2);
\draw [line width=1.5mm,-to] (36,12) -- (36,2) -- (32,2) -- (32,-2);
\draw [line width=1.5mm,-to] (30,12) -- (30,1) -- (20,1) -- (20,-6);

\draw [line width=1.5mm,-to] (18,-8) -- (21.25,-8);
\draw [line width=1.5mm,-to] (30,-4) -- (33.25,-4);

\draw [line width=1.5mm,-to] (42,0) -- (42,-6);
\draw[line width = 1.5mm, dotted] (45,10) -- (-4,10) -- (-4,14) -- (45,14) -- (45,10);
\draw[line width = 1.5mm, dotted] (45,11.5) -- (48,11.5);
\node[draw,font=\huge,text width = 3cm,align=center, fill = white] at(48,11.5) {Inputs};

\draw[line width = 1.5mm, dotted] (-4,9.5) -- (-4,-1.75) -- (25,-1.75) -- (25,9.5) --(-4,9.5);
\draw[line width = 1.5mm, dotted] (25,4) -- (48,4);
\node[draw,font=\huge,text width = 4cm,align=center, fill = white] at(48,4) {Residual};

\draw[line width = 1.5mm, dotted] (-4,-2.25) -- (-4,-10) -- (3,-10) -- (3,-2.25) --(-4,-2.25);
\draw[line width = 1.5mm, dotted] (0,-10) -- (0,-12);
\node[draw,font=\huge,text width = 4cm,align=center, fill = white] at(0,-12) {Residual restriction \\ to level 2 and 3};

\draw[line width = 1.5mm, dotted] (18,-8) -- (18,-12);
\node[draw,font=\huge,text width = 3cm,align=center, fill = white] at(18,-12) {Smoothing of level 3};

\draw[line width = 1.5mm, dotted] (24,-8) -- (24,-12);
\node[draw,font=\huge,text width =4cm,align=center, fill = white] at(24,-12) {Prolongation to level 2};

\draw[line width = 1.5mm, dotted] (30,-4) -- (30,-12);
\node[draw,font=\huge,text width = 3cm,align=center, fill = white] at(30,-12) {Smoothing of level 2};

\draw[line width = 1.5mm, dotted] (36,-4) -- (36,-12);
\node[draw,font=\huge,text width = 4cm,align=center, fill = white] at(36,-12) {Prolongation to level 1};

\draw[line width = 1.5mm, dotted] (32,0) -- (48,0);
\node[draw,font=\huge,text width = 4cm,align=center, fill = white] at(48,0) {Smoothing of level 1};

\draw[line width = 1.5mm, dotted] (44,-8) -- (48,-8);
\node[draw,font=\huge,text width = 4cm,align=center, fill = white] at(48,-8) {Output};
\node[draw,font=\huge,text width = 5cm,align=center,fill=green!80!blue] at (18,12) {$\bm{\Psi}^{(k)}$ \\ $N_x\times N_y\times N_a\times N_a\times N_f$};

\node[draw,font=\huge,text width = 5cm,align=center,fill=green!80!blue] at (12,12) {$\bm{\mu}$ \\ $N_a\times N_a\times N_f$};
\node[draw,font=\huge,text width = 5cm,align=center,fill=green!80!blue] at (24,12) {$\bm{\nu}$ \\ $N_a\times N_a\times N_f$};
\node[draw,font=\huge,text width = 5cm,align=center,fill=green!80!blue] at (6,12) {$\bm{d}$ \\ $N_x\times N_y\times N_a\times N_a\times N_f$};
\node[draw,font=\huge,text width = 5cm,align=center,fill=green!80!blue] at (0,12) {$\bm{q}$ \\ $N_x\times N_y\times N_a\times N_a\times N_f$};

\node[draw,font=\huge,text width = 5cm,align=center,fill=green!80!blue] at (30,12) {$\bm{d_3}^{\odot-1}$ \\ $\frac{N_x}{4}\times \frac{N_y}{4}\times \frac{N_a}{4}\times \frac{N_a}{4}\times N_f$};

\node[draw,font=\huge,text width = 5cm,align=center,fill=green!80!blue] at (36,12) {$\bm{d_2}^{\odot-1}$ \\ $\frac{N_x}{2}\times \frac{N_y}{2}\times \frac{N_a}{2}\times \frac{N_a}{2}\times N_f$};

\node[draw,font=\huge,text width = 5cm,align=center,fill=green!80!blue] at (42,12) {$\bm{d_1}^{\odot-1}$ \\ $N_x\times N_y\times N_a\times N_a\times N_f$};

\node[draw,font=\huge,text width = 5cm,align=center,fill=orange] at (14,8) {Multiply \\$N_x\times N_y\times N_a\times N_a\times N_f$};

\node[draw,font=\huge,text width = 5cm,align=center,fill=orange] at (6,6) {Multiply \\$N_x\times N_y\times N_a\times N_a\times N_f$};

\node[draw,font=\huge,text width = 5cm,align=center,fill=cyan] at (14,4) {$\bm{f(\;\cdot\;;\bm{w_x})}$ \\ $N_x\times N_y\times N_a\times N_a\times N_f$};

\node[draw,font=\huge,text width = 5cm,align=center,fill=orange] at (22,8) {Multiply \\ $N_x\times N_y\times N_a\times N_a\times N_f$};

\node[draw,font=\huge,text width = 5cm,align=center,fill=cyan] at (22,4) {$\bm{f(\;\cdot\;;\bm{w_y})}$ \\ $N_x\times N_y\times N_a\times N_a\times N_f$};

\node[draw,font=\huge,text width = 5cm,align=center,fill=orange] at (6,0) {Add \\ $N_x\times N_y\times N_a\times N_a\times N_f$};

\node[draw,font=\huge,text width = 5cm,align=center,fill=orange] at (0,0) {Subtract\\ $N_x\times N_y\times N_a\times N_a\times N_f$};

\node[draw,font=\huge,text width = 5cm,align=center,fill=lightgray] at (42,-8) {$\bm{\Psi}^{(k+1)}$  \\ $N_x\times N_y\times N_a\times N_a\times N_f$};

\node[draw,font=\huge,text width = 5cm,align=center,fill=cyan] at (0,-4) {$\bm{f(\;\cdot\;;\bm{w_{R}})}$  \\ $\frac{N_x}{2}\times \frac{N_y}{2} \times \frac{N_a}{2}\times \frac{N_a}{2}\times N_f$};

\node[draw,font=\huge,text width = 5cm,align=center,fill=cyan] at (0,-8) {$\bm{f(\;\cdot\;;\bm{w_{R}})}$  \\ $\frac{N_x}{4}\times \frac{N_y}{4} \times \frac{N_a}{4}\times \frac{N_a}{4}\times N_f$};

\node[draw,font=\huge,text width = 5cm,align=center,fill=yellow] at (18,-8) {$\text{J}(\;\cdot\;)$  \\ $\frac{N_x}{4}\times \frac{N_y}{4} \times \frac{N_a}{4}\times \frac{N_a}{4}\times N_f$};

\node[draw,font=\huge,text width = 5cm,align=center,fill=yellow] at (30,-4) {$\text{J}(\;\cdot\;)$   \\ $\frac{N_x}{2}\times \frac{N_y}{2} \times\frac{N_a}{2}\times \frac{N_a}{2}\times N_f$};

\node[draw,font=\huge,text width = 5cm,align=center,fill=yellow] at (42,0) {$\text{J}(\;\cdot\;)$   \\ $N_x\times N_y\times N_a\times N_a\times N_f$};

\node[draw,font=\huge,text width = 5cm,align=center,fill=teal!60!] at (36,-4) {UpSamp  \\ $N_x\times N_y\times N_a\times N_a\times N_f$};

\node[draw,font=\huge,text width = 5cm,align=center,fill=teal!60!] at (24,-8) {UpSamp  \\ $\frac{N_x}{2}\times \frac{N_y}{2} \times \frac{N_a}{2}\times \frac{N_a}{2}\times N_f$};

\end{tikzpicture}}
\caption{Multigrid network, $\text{MG}$, representing a single multigrid iteration. This network performs a single multigrid iteration on the flux of a single energy group. It takes the angular flux ($\bm{\Psi}^{(k)}$), source ($\bm{q}$), angular velocities ($\bm{\mu}$,$\bm{\nu}$) and the strictly diagonal coefficients ($\bm{d}_{\bm{1}}$), along with the coarser resolution coefficients, as inputs (green boxes). A number of layer operations are performed, mathematical operations in orange, convolutional operations in cyan, sub-model operations in yellow and upsampling in teal. The sub-models can be iterated on multiple times. Finally, it outputs the flux of the next multigrid iteration flux ($\bm{\Psi}^{(k+1)}$). Arrow origins show which layer the data originated and the end of the arrow shows which layer takes that data as input. Dimensions of layers are given on the second line of each box.}
\label{fig:multi_grid_net}
\end{figure}

The residual calculation using convolutional layers is given by Equation~\eqref{Petrov} using the Petrov-Galerkin method. 
The restriction is done in two stages. In the first stage the residual is modified by multiplying the flux in each angle by its weight, $p_n$,  and $\Delta x \Delta y$:
\begin{equation}
    \bm{r}_{\bm{1},n}^{mod}  = p_n \bm{r}_{\bm{1},n} \Delta x\Delta y, 
\end{equation}
in which $k$ represents the iteration level (it increments after a multi-grid cycle such as shown in Figure~\ref{fig:multi_grid_it})
and in the second stage a convolutional operation 
is performed:
\begin{equation}
    \bm{r}^{(k)}_{\bm{2}} = \bm{f}(\bm{r}^{mod,(k)}_{\bm{1}};\bm{w_{R}}),
\end{equation}
with filter weights:
\begin{gather}
\bm{w_{R}}
 =
 \begin{bmatrix}
 1 &  1 \\
1 &  1 \\
 \end{bmatrix}.
\end{gather}
An upsampling convolutional layer is applied from one solution level to the next finest level using the convolutional operation $\text{UpSamp}(\;\cdot\;)$, see Figure~\ref{fig:multi_grid_it}. This operation simply copies the coarser grid solution to the finner grid solutions. Thus, upsampling layers repeat values within an array, increasing the dimensions of the data~\cite{chollet2015keras}, resulting in an approximation for the data at a finer mesh: 
\begin{equation}
    \widetilde{\bm{\Psi_1}}^{(k)} = \text{UpSamp}( \bm{\Psi}_2^{(k)}) 
\end{equation}
Although not shown in Figure~\ref{fig:multi_grid_net} the Jacobi networks, given by the yellow boxes with a $\text{J}(\;\cdot\;)$, also have a set of $\mu_n$ and $\nu_n$ as inputs, which is dependent upon the restriction level in angle.

Figure~\ref{fig:multi_grid_net} shows how the multigrid iteration would look as a single network. Green boxes contain the inputs, blue boxes are convolutional layers, orange boxes are mathematical functions as layers, yellow boxes are sub-networks, teal boxes are upsampling layers and the grey box is the output of the network. The second line in each box is the dimension of the output.  This can be written as:
\begin{equation}
    \bm{\Psi}_{\bm{1}}^{(k+1)} = \text{MG}\left(\bm{\Psi}_{\bm{1}}^{(k)},\bm{q}_{\bm{1}},\bm{a}_{\bm{1}},\bm{\mu},\bm{\nu},\bm{d},\bm{d}_{\bm{1}}^{\odot-1},\bm{d}_{\bm{2}}^{\odot-1},\bm{d}_{\bm{3}}^{\odot-1}\right)
\end{equation}
where $\text{MG}$ is a function that calculates the result of one sawtooth multigrid iteration.
%\clearpage

\subsection{Optimised Anisotropic Non-Linear
Petrov-Galerkin Dissipation}  
%\subsection{The Non-linear Petrov-Galerkin Method}
\label{sec:petrov_galerkin}

%The diffusion based discrete residual for a given direction $n$ is given by:
%\begin{equation}\label{eq:diff_res}
%   \bm{ R_n}= \nabla \alpha_{on}\bm{o_n}\bm{o_n^T}\nabla\Psi_n,
%\end{equation}
%where:
% $m_i$ is the mass given by $dx dy$ and:
%\begin{equation}
%    \alpha_{on} = \sqrt{a_{xn}^2 + a_{yn}^2},
%\end{equation}
%where: 
%\begin{equation}
%    a_{xn} = \frac{\mu_n}{dx}\text{ , } a_{yn} = \frac{\nu_n}{dy},
%\end{equation}
%and:
%\begin{equation}
%    \bm{o_n} = \frac{a_{xn}}{\alpha_{on}},\frac{a_{yn}}%{\alpha_{on}}.
%\end{equation}.
%Equation~\ref{eq:diff_res} is equal to:
%\begin{equation}
%    \bm{R_n}= \left[\frac{a_{xn^2}}{a_{on}}\frac{\delta^2\bm{\Psi_n}}{\delta x^2} + \frac{a_{xn}a_{yn}}{a_{on}}\frac{\frac{\delta(\delta\bm{\Psi_n})}{\delta y}}{\delta x}  +
%    \frac{a_{yn}a_{xn}}{a_{oi}}\frac{\frac{\delta(\delta\bm{\Psi_n})}{\delta x}}{\delta y} 
%    +\frac{a_{yn^2}}{a_{on}}\frac{\delta^2\bm{\Psi_n}}{\delta y^2}\right]
%\end{equation}

%\par\noindent
%Define ${C_x}_i = (A_x C)|_i$ , 
%${C_y}_i = (A_y C)|_i$  . 

If $\bm{\Psi}_{\bm{x}n,g} =  \bm{f}(\bm{\Psi}_{n,g};\bm{w_{x}}) $ and $\bm{\Psi}_{\bm{y}n,g} =  \bm{f}(\bm{\Psi}_{n,g};\bm{w_{y}})$ then a diffusion coefficient, based on the Petrov-Galerkin method in \cite{CODINA1993325} and also presented in \cite{donea2003finite}, can be given by:
\begin{equation}
   \bm{ k}_{\bm{abs}\,n,g} = \frac{\alpha_{kabs}|\bm{R}_{n,g}| h  }{\epsilon_k+\frac{1}{\cal D}(|\bm{\Psi}_{\bm{x}{n,g}}|+|\bm{\Psi}_{\bm{y}{n,g}}|)} \; ,
\end{equation}
in which the dimensionality of the system here is ${\cal D}=2$. 
A second diffusion coefficient, based on \cite{HUGHES1986} and also described in \cite{FANG2013540,donea2003finite} all be it with the use of the 1-norm rather than 2-norms,  can be given by:
\begin{equation}
    \bm{k}_{\bm{square}{\, n,g}} = \frac{\alpha_{ksquare} {\bm{R}_{n,g}}^2h}{\epsilon_k+\frac{1}{\cal D}(|\bm{q}_{n,g}\odot\bm{\Psi}_{\bm{x}{n,g}}|+|\bm{q}_{n,g}\odot\bm{\Psi}_{\bm{y}{n,g}}|)(\bm{\Psi}_{\bm{x}{n,g}}^2+\bm{\Psi}_{\bm{y}{n,g}}^2)} \; ,
\end{equation}
where $\bm{q}_{n,g}$ is:
\begin{equation}
    \bm{q}_{n,g} = \frac{\mu_n\bm{\Psi}_{\bm{x}n,g} +\nu_n\bm{\Psi}_{\bm{y}n,g} }{\epsilon_k+{\bm{\Psi}_{\bm{x}n,g}}^2+{\bm{\Psi}_{\bm{y}n,g}}^2 } \;.
\end{equation}
For both diffusion coefficients, $h= \frac{1}{2}(\Delta x + \Delta y)$ and $\epsilon_k$ is a small value chosen so that the resulting diffusion does not become too large --- it is set to $\epsilon_k=0.001$ in this work. 
The $\mathcal{O}(1)$ scaling coefficients $\alpha_{kabs}$, $\alpha_{ksquare}$ are defined here as:
\begin{equation}
    \alpha_{kabs} =\frac{2^p}{16} , \qquad
    \alpha_{ksquare} =\frac{2^p}{2},
    \label{scalars_alpha}
\end{equation}
where $p$ is the polynomial order of the finite element expansion: $p=1$ for linear $3\times3$ filters, $p=2$ for quadratic $5\times5$ filters and $p=3$ for cubic $7\times7$ filters.  The scalars, 
Equations~\eqref{scalars_alpha}, increase with increasing $p$
because the residual $\bm{R}_{n,g}$ gets smaller with increasing $p$. If $p>1$ then $\bm{R}_{n,g}$ can instead be determined by subtracting the lower order expansion from the higher order expansion: 
\begin{equation}
    \bm{R}_{n,g} =  \left[\mu_n 
     \bm{f}(\bm{\Psi}_{n,g};\bm{w_{x}^{}}) +\nu_n \bm{f}(\bm{\Psi}_{n,g};\bm{w_{y}^{}})\right] -  \left[\mu_n 
     \bm{f}(\bm{\Psi}_{n,g};\bm{w_{x}^{Low}}) +\nu_n \bm{f}(\bm{\Psi}_{n,g};\bm{w_{y}^{Low}})\right]
\end{equation}
in which $\bm{w_{x}^{Low}}$, $\bm{w_{y}^{Low}}$ are lower order (by one) filters
than those used in the rest of the simulation e.g. if the filter size in 3D is $5\times5\times5$ then the lower order size
might be $3\times3\times3$. This $\bm{R}_{n,g}$ effectively forms the residual of the discrete system of equations and results in a
particularly impressive performing method. To obtain high rates of convergence use: 
\begin{equation}
    \bm{R}_{n,g} =  \left[\mu_n 
     \bm{f}(\bm{\Psi}_{n,g};\bm{w_{x}^{High}}) +\nu_n \bm{f}(\bm{\Psi}_{n,g};\bm{w_{y}^{High}})\right] -  \left[\mu_n 
     \bm{f}(\bm{\Psi}_{n,g};\bm{w_{x}^{}}) +\nu_n \bm{f}(\bm{\Psi}_{n,g};\bm{w_{y}^{}})\right] 
     \label{R^High} 
\end{equation}
in which $\bm{w_{x}^{High}}$, $\bm{w_{y}^{High}}$ are higher order (by one) filters. 
Equation~\eqref{R^High} can be derived by placing the current solution 
$\bm{\Psi}_{n,g}$ into the higher order discretised equations and then subtract the current discretised equations (which have a zero residual) from the high order discretised equations (which have a non-zero residual). 
The result is Equation~\eqref{R^High}.  
A mixed mass approach to residual approximation can be similarly derived to obtain the residual approximation:
\begin{equation}
    %\bm{R}_{n,g} =  \beta ( {M_L}^{-1} M - I) 
    %\left(\mu_n \bm{f}(\bm{\Psi_n};\bm{w_{x}}) +\nu_n \bm{f}(\bm{\Psi_n};\bm{w_{y}}) \right),
    \bm{R}_{n,g} =  \beta_r 
    \bm{f}( \; \mu_n \bm{f}(\bm{\Psi}_{n,g};\bm{w_{x}}) +\nu_n \bm{f}(\bm{\Psi}_{n,g};\bm{w_{y}}) ; \; {m_L}^{-1} \bm{m} - \bm{I} ),
\end{equation}
where $m_L=\Delta x\Delta y$ is the lumped mass term and thus $m_L^{-1}=\frac{1}{\Delta x\Delta y}$, $\bm{m}$ is the mass filter associated with the finite element matrices, see appendix. $\beta_r$ is a coefficient where $3$ has been found effective. A maximum diffusion coefficient is considered to be the diffusion whose magnitude is the same as that of advection and is thus:
\begin{equation}
    \bm{k}_{\bm{max}n,g} = \sqrt{(\Delta x \mu_n)^2 + (\Delta y\nu_n)^2} \; , 
\end{equation}
and the final diffusion coefficient, combining the above conservative values of diffusion, is given by:
\begin{equation}
     \bm{k}_{n,g} = \bm{min}\{ \bm{k}_{\bm{max}n,g}, \bm{k}_{\bm{abs}\, n,g}, \bm{k}_{\bm{square}\, n,g}\}. 
\end{equation}

\textit{Anisotropic Residual based Diffusion for Non-Linear Petrov-Galerkin Dissipation}
\par\noindent 
A major advanatage of representing the discrete equation residual from just the gradients is that we can form residual contributions from 
different directions and in this way form anisotropic diffusion coefficients that stabilize the method. This is what we do here and is the approach used in the applications. 
Now the residual in each direction in
tern can also be estimated from the mixed mass approach as: 
\begin{eqnarray}
    \bm{R}_{\bm{x}n,g} &= & \alpha_r \mu_n 
    \bm{f}( \; \bm{f}(\bm{\Psi}_{n,g};\bm{w_{x}})  ; \; {m_L}^{-1} \bm{m} - \bm{I} ), \\
    \bm{R}_{\bm{y}n,g} &= & \alpha_r \nu_n
    \bm{f}( \;  \bm{f}(\bm{\Psi}_{n,g};\bm{w_{y}}) ; \; {m_L}^{-1} \bm{m} - \bm{I} ).
\end{eqnarray}
This provides an opportunity to apply anisotropically the Petrov-Galerkin diffusion – that is to apply different
diffusion in different directions. In its simplest form this can be achieved through for example (using the
previous approach but for each coordinate in turn): 
\begin{eqnarray}
\bm{k}_{\bm{xabs}\, n,g} = \frac{ \alpha_{kabs} \vert \bm{R}_{\bm{x}n,g} \vert \Delta x  } 
{(\epsilon_k +\vert \bm{f}(\bm{\Psi}_{n,g};\bm{w_{x}}) \vert )}, \;\;\; 
\bm{k}_{\bm{yabs}\, n,g} = \frac{ \alpha_{kabs} \vert \bm{R}_{\bm{y}n,g} \vert \Delta y  } 
{(\epsilon_k +\vert \bm{f}(\bm{\Psi}_{n,g};\bm{w_{y}}) \vert )},
\end{eqnarray}
%and becomes simply: 
%\begin{eqnarray}
%\bm{k_{xabs}}_{\, n,g} = \alpha_{kabs} \vert \mu_n\vert \Delta x, \;\;\; 
%\bm{k_{yabs}}_{\, n,g} = \alpha_{kabs} \vert \nu_n\vert \Delta y,
%\end{eqnarray}
and the second set of diffusion coefficients are:
%the same become:
\begin{equation}
    \bm{k}_{\bm{x\, square}\, n,g} = \frac{\alpha_{ksquare}{ {\bm{R}_{\bm{x}n,g}}^2}\Delta x}{\epsilon_k+  \vert \mu_n \vert {\bm{\Psi}_{\bm{x}n,g}}^2 }.  \; \;\; 
    \bm{k}_{\bm{y\, square}\, n,g} = \frac{\alpha_{ksquare}{ {\bm{R}_{\bm{y}n,g}}^2}\Delta y}{\epsilon_k+  \vert \nu_n \vert {\bm{\Psi}_{\bm{y}n,g}}^2 },
\end{equation}
and the diffusion coefficients in the x- and y-directions respectively are:
\begin{equation}
     \bm{k}_{\bm{x}n,g} = \bm{min}\{ \bm{k}_{\bm{x\,max}n,g}, \bm{k}_{\bm{{x\,abs}}\, n,g}, \bm{k}_{\bm{x\, square}\, n,g}\}, \;\;\; 
     \bm{k}_{\bm{y}n,g} = \bm{min}\{ \bm{k}_{\bm{y\, max}n,g}, \bm{k}_{\bm{y\,abs}\, n,g}, \bm{k}_{\bm{y\, square}\, n,g}\},
     \label{ano-k}
\end{equation}
in which $\bm{k}_{\bm{x\,max}n,g}= \Delta x \vert \mu_n \vert$ and  $\bm{k}_{\bm{y\,max}n,g}=  \Delta y \vert \nu_n \vert$.
This form, Equation~\eqref{ano-k},  of anisotropic diffusion is the approach followed here because on structured grids aligned with the diffusion directions this is particularly effective with minimal discretisation error. 
However, Equation~\eqref{ano-k} can be generalised further by avoiding assuming that the diffusion is aligned with the coordinate
system. Initially, for simplicity, we will assume the system is 2D. This then enables us to define another 
residual from: 
\begin{equation}
    \bm{R}_{\bm{xy}n,g} =  \alpha_r  
    \bm{f}( \; \mu_n \bm{f}(\bm{\Psi}_{n,g};\bm{w_{x}}) +\nu_n \bm{f}(\bm{\Psi}_{n,g};\bm{w_{y}}) ; \; {m_L}^{-1} \bm{m} - \bm{I} ),
\label{R_xy}
\end{equation}
with an associated diffusion coefficient of 
\begin{eqnarray}
\bm{k}_{\bm{xyabs}\, n,g} = 
\frac{ \alpha_{kabs} \vert \bm{R}_{\bm{xy}n,g} \vert }
{(\epsilon_k +\frac{\vert \bm{f}(\bm{\Psi}_{n,g};\bm{w_{x}}) \vert}{{\cal D}\Delta x  }
+\frac{\vert \bm{f}(\bm{\Psi}_{n,g};\bm{w_{y}}) \vert}{{\cal D}\Delta y} ) }. 
\end{eqnarray}
Now we have three diffusion coefficients so one might find the largest of these three  
coefficients and rotate the coordinate system so that it is aligned with this diffusion direction, either $(\mu_n ,0)^T$,  $(0, \nu_n )^T$  
then project the other two directions so they are orthogonal to this direction and
choose the largest of the remaining diffusion coefficients (after the projection). This becomes the orthogonal
diffusion then rotate the system back to the original system to form the final diffusion tensor. A similar
approach can be applied in 3D but then 6 residuals are formed e.g. 
\begin{equation}%{a_b}_c
    \bm{R}_{\bm{xyz}n,g} =  \alpha_r  
    \bm{f}( \; \mu_n \bm{f}(\bm{\Psi}_{n,g};\bm{w_{x}}) +\nu_n \bm{f}(\bm{\Psi}_{n,g};\bm{w_{y}}) +\xi_n \bm{f}(\bm{\Psi}_{n,g};\bm{w_{z}}) ; \; {m_L}^{-1} \bm{m} - \bm{I} ),
\label{R_xyz}
\end{equation}
as
well as $\bm{R}_{\bm{xy}n,g}$ as defined above in Equation~\eqref{R_xy} and where $\xi_n$ is the 3$^{\text rd}$ direction on the unit sphere in the discrete ordinate angular discretisation.

\subsection{Neural Network Filters for the Convolutional Finite Element Method (ConvFEM)}

\begin{figure}[H]
\centering
\begin{minipage}{.45\textwidth}
  \centering
    \includegraphics[scale=0.5]{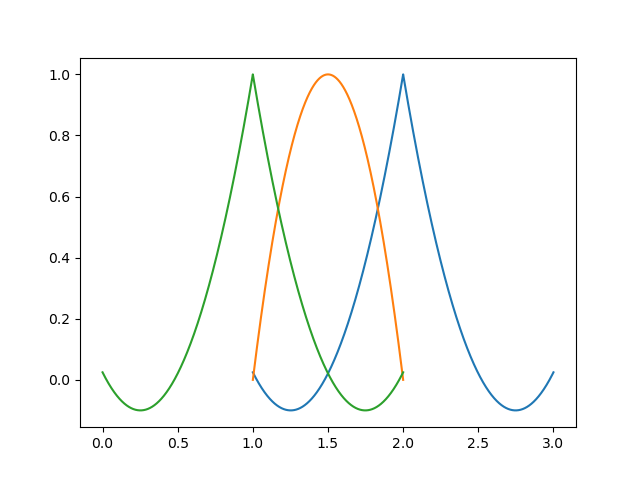}
  \subcaption{Quadratic basis functions for a single element}
  \label{fig:basis_functions_1}
\end{minipage}%
\hfill
\begin{minipage}{.45\textwidth}
  \centering
  \includegraphics[scale=0.5]{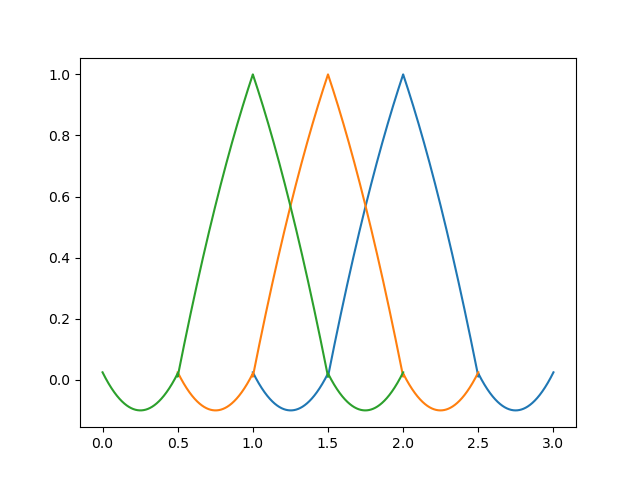}
  \subcaption{Averaged basis functions for three elements}
  \label{fig:basis_functions_3}
\end{minipage}
\caption{FEM Basis functions for quadratic 1D elements.}
\label{fig:basis_functions}
\end{figure}

The basic idea is to add discretisation stencils together in order to form an average stencil that looks the same at every node of the FEM mesh. 
One can see the stencils are different by looking at the 1D quadratic element. For this element, the node at the centre of the element has a direct link (through the FEM stencil) only to the nodes of that element (3 of them including the node itself) while the other nodes have direct links to all the nodes of the 2 elements they belong too and are thus linked directly to 5 nodes including the current node. This difference in stencils can also be seen from node to
node in Figure \ref{fig:basis_functions_1}. 
The basis functions for quadratic 1D element 
discretisations are shown in figure~\ref{fig:basis_functions_1}. 
To form the new ConvFEM discretisation, these different discretisations are simply added together to form another discretisation which looks the same everywhere. If there is a lumped basis
function then it looks like an average of the two basis functions – see Figure \ref{fig:basis_functions_3}. 

\begin{figure}[H]
\centering
\begin{minipage}{.45\textwidth}
  \centering
  \scalebox{0.1}{\begin{tikzpicture}
\draw[step=15.0,black,thin] (0,0) grid (60,60);
\foreach \x in {0,15,30,...,60}
\foreach \y in {0,15,30,...,60}
\draw[black, fill] (\x,\y) circle (20pt); 
\draw[green,thick] (-7.5,-7.5) -- (-7.5,22.5) -- (22.5,22.5) --  (22.5,-7.5) -- (-7.5,-7.5)  ;
\foreach \x in {0,15,30}
\foreach \y in {0,15,30}
\draw[blue, fill] (\x,\y) circle (40pt); 

\draw[blue] (0,0) circle (70pt); 
\end{tikzpicture}}
  \subcaption{Quadratic element indicating in the green box the nodal equations that are to be averaged. }
  \label{fig:quad}
\end{minipage}%
\hfill
\begin{minipage}{.45\textwidth}
  \centering
 \scalebox{0.1}{\begin{tikzpicture}
\draw[step=15.0,black,thin] (0,0) grid (60,60);
\foreach \x in {0,15,30,...,60}
\foreach \y in {0,15,30,...,60}
\draw[black, fill] (\x,\y) circle (20pt); 
\draw[green,thick] (-7.5,-7.5) -- (-7.5,37.5) -- (37.5,37.5) --  (37.5,-7.5) -- (-7.5,-7.5)  ;
\foreach \x in {0,15,30,45}
\foreach \y in {0,15,30,45}
\draw[blue, fill] (\x,\y) circle (40pt); 

\draw[blue] (0,0) circle (70pt); 
\end{tikzpicture}}
  \subcaption{Cubic element indicating in the green box the nodal equations that are to be averaged. }
  \label{fig:cubic}
\end{minipage}
\caption{Nodes within green boxes will have their associated discrete equations averaged. The blue nodes indicate the nodes of the 2D quadratic and cubic elements. }
\label{fig:av_ele}
\end{figure}
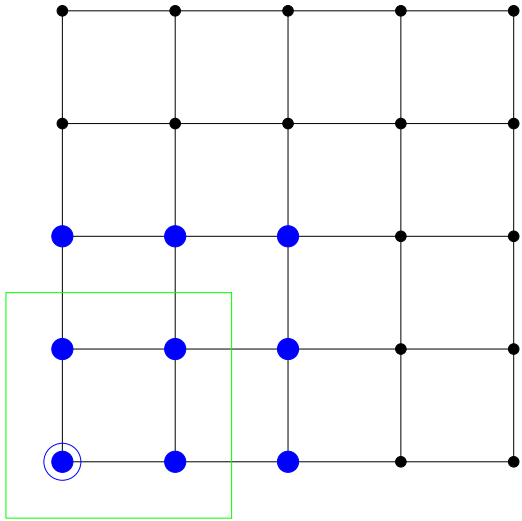
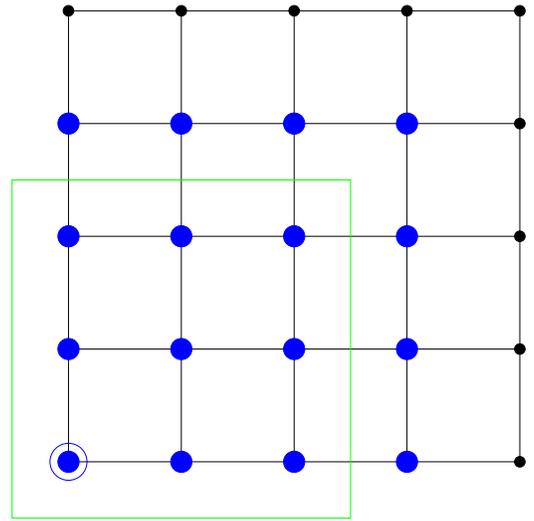

For a 2D quadratic element (Figure \ref{fig:quad}) the node in the centre of the element is connected directly to only the nodes of the element (8 of them) and the edge nodes of the element are directly connected only to the 2 elements that it belongs to (15-1=14 of them) and the corner nodes are connected to all of the nodes belonging to the four elements that it belongs to (25-1=24 of them).  At the bottom we have highlighted (green box of Figure \ref{fig:quad}) the 4
nodes around which the stencils need to be averaged for 2D and for
3D quadratic elements there are similarly 8 nodes around which averaging is necessary. For cubic elements, one needs to average, in 2D, over the 9 nodes shown in Figure \ref{fig:cubic} with the green box around these 9 nodes. 

For example, when adding the two equations together for 1D quadratic elements the mass discretisation becomes: 
\begin{eqnarray}
\int_E \frac{1}{2} (N_i \sum_j N_j C_j + S_i\sum_j S_j C_j)dV 
\end{eqnarray}
in which $N_i$ and $S_i$ 
are the basis functions of the two discretisations
in 1D and for diffusion discretisation we thus have: 
\begin{eqnarray}
\int_E \frac{1}{2} \left(\frac{\partial N_i}{\partial x} \sum_j \frac{\partial N_j}{\partial x} C_j + 
\frac{\partial S_i}{\partial x} \sum_j \frac{\partial S_j}{\partial x} C_j \right)dV.  
\end{eqnarray}

The finite element method (FEM) filter weights used in section~\ref{sec:petrov_galerkin} are constructed from the finite element basis functions. The basis functions for a number of nodes, dependent on the order, are summed together and averaged. The 1D quadratic basis functions are shown in Figure~\ref{fig:basis_functions_1}. By taking the basis functions for the two neighbouring nodes, and averaging them, the overall basis functions for the same element are shown in Figure~\ref{fig:basis_functions_3}. 
See the appendix for all the 2D filters needed for the ConvFEM discretisation used here. 

The required elements needing to be averaged for quadratic and cubic elements are shown in Figure~\ref{fig:av_ele}. 
% The residual with the stabilisation term used within the rest of the multigrid solution is therefore:
% \begin{equation}
%     r = s - A\phi + r_{stab},
% \end{equation}

% \begin{equation}
%     r_i = s_i - \sigma \psi_i - \Omega \nabla \psi_i + \nabla k\nabla \psi_i
% \end{equation}
% \begin{equation}
%     \Omega \nabla \psi_i = u_i \frac{\delta \psi_i}{\delta x} +  v_i \frac{\delta \psi_i}{\delta y}
% \end{equation}
% Upwind transport term ( Assuming dx,dy =1 and u,v $>$ 0):
% \begin{equation}
%     (u \psi_{i,j} - u_i \psi_{i-1,j}) + (v \psi_{i,j} - v \psi_{i,j-1})
% \end{equation}
% Petrov Galerkin (with same assumptions):
% \begin{equation}
% \begin{aligned}
%     \frac{1}{2}(u \psi_{i+1,j} - u_i \psi_{i-1,j}) +  \frac{1}{2}(v \psi_{i,j+1} - v \psi_{i,j-1})+ \\\frac{1}{2}(-(k_{i+1,j}+k_{i,j})(\psi_{i+1,j})-(k_{i,j+1}+k_{i,j})(\psi_{i,j+1})+\\(4k_{i,j}+k_{i+1,j}+k_{i-1,j}+k_{i,j-1}+k_{i,j+1})(\psi_{i,j})-\\\frac{1}{2}(k_{i-1,j}+k_{i,j})(\psi_{i-1,j}-\frac{1}{2}(k_{i,j-1}+k_{i,j})(\psi_{i,j-1}))
%     \end{aligned}
% \end{equation}
% The diffusion for an element $i$ is therefore  given by:
% \begin{equation}\label{eq:conv_filter_version}
%   d_i =  k_i\bm{G}(\bm{\Psi})|_{i}+\bm{G}(\bm{K}\odot\bm{\Psi})|_{i} - \psi_i\bm{G}(\bm{K})|_{i},
% \end{equation}
% where $\bm{K}$ is a matrix containing every $k_i$ element and $\bm{G}$ is the diffusion filter used. The residual for element $i$ is therefore given as:
% \begin{equation}\label{eq:disc_ang_hfm}
%     r_i = q_{i} - \left[ \mu_n\frac{\delta}{\delta x} + \nu_n\frac{\delta}{\delta y} + \sigma_{i} \right ]\psi_{i} - d_i.
% \end{equation}
\section{Results}\label{sec:results}
\subsection{Straight Duct problem}
%- upwind method}
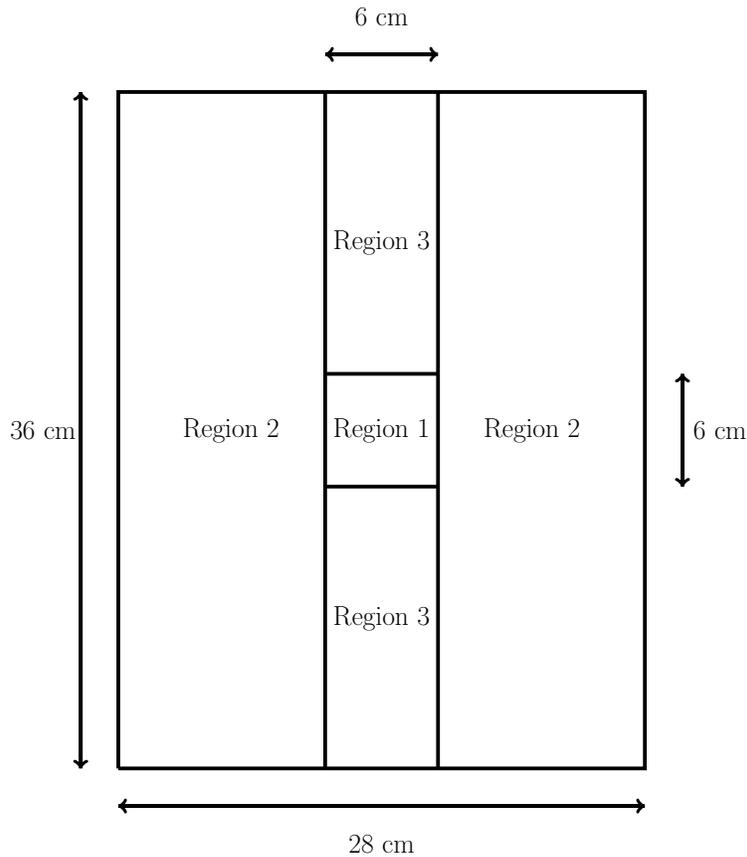
\begin{figure}[H]
\centering
\scalebox{0.5}{\begin{tikzpicture}

\draw [line width=1mm] (0,0) -- (14,0) -- (14,18) -- (0,18) -- (0,0);
\draw [line width=1mm,-to] (0,-1) -- (14,-1);
\draw [line width=1mm,-to] (14,-1) -- (0,-1);
\node[font=\huge,text width = 3cm,align=center] at (7,-2) {28 cm};
\draw [line width=1mm,-to] (-1,0) -- (-1,18);
\draw [line width=1mm,-to](-1,18) --  (-1,0);
\node[font=\huge,text width = 3cm,align=center] at (-2,9) {36 cm};

\draw [line width=1mm] (5.5,0) -- (5.5,18);
\draw [line width=1mm] (8.5,0) -- (8.5,18);
\draw [line width=1mm] (8.5,7.5) -- (5.5,7.5);
\draw [line width=1mm] (8.5,10.5) -- (5.5,10.5);

\node[font=\huge,text width = 3cm,align=center] at (7,9) {Region 1};
\node[font=\huge,text width = 3cm,align=center] at (3,9) {Region 2};
\node[font=\huge,text width = 3cm,align=center] at (11,9) {Region 2};

\node[font=\huge,text width = 3cm,align=center] at (7,4) {Region 3};
\node[font=\huge,text width = 3cm,align=center] at (7,14) {Region 3};

\draw [line width=1mm,-to] (5.5,19) -- (8.5,19);
\draw [line width=1mm,-to](8.5,19) -- (5.5,19);
\node[font=\huge,text width = 3cm,align=center] at (7,20) {6 cm};
\draw [line width=1mm,-to] (15,7.5) -- (15,10.5);
\draw [line width=1mm,-to](15,10.5) -- (15,7.5);
\node[font=\huge,text width = 3cm,align=center] at (16,9) {6 cm};
\end{tikzpicture}}
\caption{Diagram of single duct problem.}
\label{fig:sing_duct_mesh}
\end{figure}
\begin{table}[h!]
\centering
 \begin{tabular}{l c c c} 
 \toprule
       & Source(\si{neutrons.cm^{-2}.s^{-1}})& $\Sigma^a$(\si{\per\cm}) & $\Sigma^s$ (\si{\per\cm}). \\[0.5ex] 

 Region 1 & 1 & 0.5 & 0 \\ 
 %\hline
 Region 2 & 0 & 0.5 & 0 \\ 
  Region 3 & 0 & 0 & 0 \\ \\[0.5ex]  

\end{tabular}
\caption{Source and cross sections for the three regions of the straight duct problem.}
\label{tab:sing_duct_data}
\end{table}

Figure~\ref{fig:sing_duct_mesh} shows the diagram of the single duct problem. This domain is a $36\si{cm} \times 28\si{cm}$ and contains a $6\si{cm} \times 6\si{cm}$ source in the centre. Two regions of absorbing material are either side of the source and two ducts of width $6\si{cm}$ run along the y-axis either side of the source. The source and cross sections for the three regions are listed in table~\ref{tab:sing_duct_data}. This problem is mono-energetic, so no scattering occurs, and has a source, so no power iteration is required. Four different spatial resolutions are used: $45 \times 35$, $90 \times 70$, $180 \times 140$ and $360 \times 280$. These four spatial resolutions have spacing between the nodes or cells of $\Delta x=0.8\si{cm}$, $\Delta x=0.4 \si{cm}$, $\Delta x=0.2\si{cm}$ and $\Delta x=0.1\si{cm}$ respectively all with $\Delta y=\Delta x$. All dimensions of length are in $\si{cm}$. Simulations with $\Delta x=0.8, 0.4, 0.2 \text{ and } 0.1$ are resolved with $N_f=8$ and $N_a=4$ resulting in $128$ total angular directions. For the finer resolution in angle simulations use  $N_f=8$ and $N_a=8$ resulting in $512$ total angular directions and $\Delta x =0.1$. The finer resolution solutions were generated using 500 multigrid iterations and all other results in this section were generated using 100 multigrid iterations. 
%In all cases $\Delta x=\Delta y$. 

  \begin{figure}[H]
\centering
\begin{minipage}{.45\textwidth}
  \centering
  \includegraphics[scale=0.5]{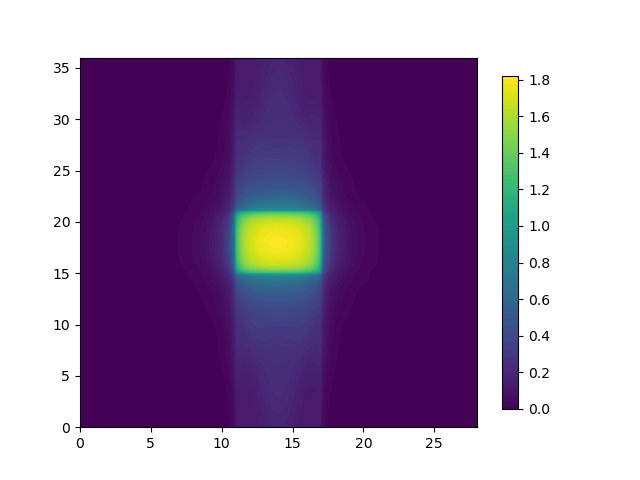}
  \subcaption{Scalar flux (\si{neutrons.cm^{-2}.s^{-1}}) across the straight duct problem.}
  \label{fig:sing_duct_up_2d}
\end{minipage}%
\hfill
\begin{minipage}{.45\textwidth}
  \centering
  \includegraphics[scale=0.5]{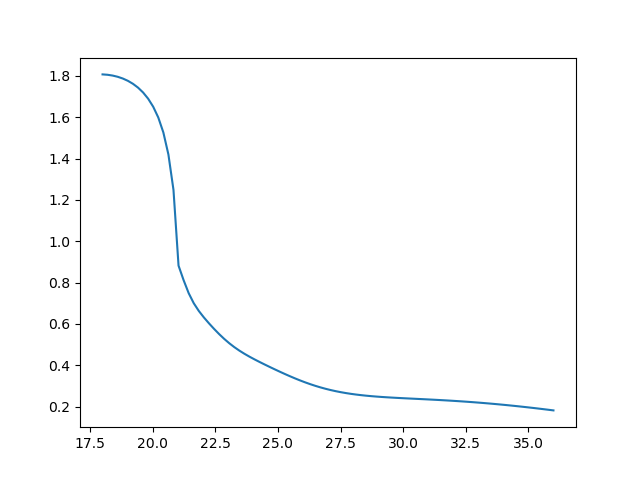}
  \subcaption{Scalar flux (\si{neutrons.cm^{-2}.s^{-1}}) vs y (\si{cm}) at x =14 \si{cm}.}
  \label{fig:sing_duct_up_1d}
\end{minipage}
\caption{Solution for the straight duct problem generated using a neural network with a pure upwind solver with a spatial resolution of $\Delta x = 0.2\si{cm}$.}
\label{fig:sing_duct_up}
\end{figure}
Figure~\ref{fig:sing_duct_up} shows the solution for the single duct problem with the upwind method. 
Notice that there is a sharper change in flux at the boundary between source and tunnel, at y = 21\si{cm}. The centre of this problem is at x = 14\si{cm} and y = 18cm, which appears at the left-hand side of figure~\ref{fig:sing_duct_up}. 

%\subsection{Straight Duct problem - Petrov-Galerkin}

%   \begin{figure}[H]
% \centering
% \begin{minipage}{.45\textwidth}
%   \centering
%   \includegraphics[scale=0.5]{figures/sing_duct_Pg_3/flux_2d.png}
%   \subcaption{Scalar flux (\si{neutrons.cm^{-2}.s^{-1}}) across the straight duct problem with the Petrov-Galerkin.}
%   \label{fig:fa_flux}
% \end{minipage}%
% \hfill
% \begin{minipage}{.45\textwidth}
%   \centering
%   \includegraphics[scale=0.5]{figures/sing_duct_Pg_3/flux_1d.png}
%   \subcaption{Scalar flux (\si{neutrons.cm^{-2}.s^{-1}}) vs y (\si{cm}) at x =70 \si{cm}.}
%   \label{fig:fa_keff}
% \end{minipage}
% \caption{Solution for the straight duct problem with the Petrov-Galerkin using Linear FEM filters.}
% \label{fig:pg_lin}
% \end{figure}

  \begin{figure}[H]
\centering
\begin{minipage}{.45\textwidth}
  \centering
  \includegraphics[scale=0.5]{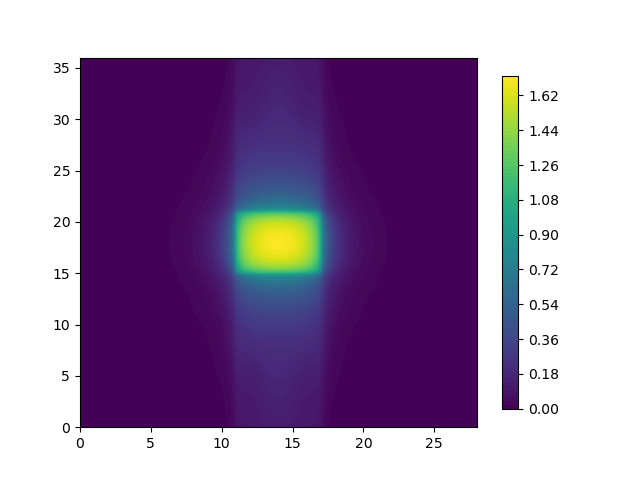}
  \subcaption{Scalar flux (\si{neutrons.cm^{-2}.s^{-1}}) across the straight duct problem with the Petrov-Galerkin.}
  \label{fig:pg_quad_2d}
\end{minipage}%
\hfill
\begin{minipage}{.45\textwidth}
  \centering
  \includegraphics[scale=0.5]{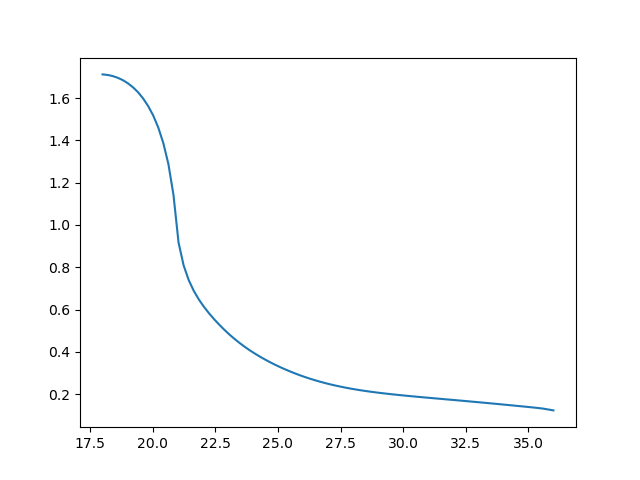}
  \subcaption{Scalar flux (\si{neutrons.cm^{-2}.s^{-1}}) vs y (\si{cm}) at x =14 \si{cm}.}
  \label{fig:pg_quad_1d}
\end{minipage}
\caption{Solution for the straight duct problem with the Petrov-Galerkin generated using a neural network with Quintic ConvFEM filters with a spatial resolution of $\Delta x = 0.2\si{cm}$ and using 128 angular directions.}
\label{fig:pg_quad}
\end{figure}

Figure~\ref{fig:pg_quad} shows the solution for the single duct problem with Petrov-Galerkin residual added and using quadratic ConvFEM filters. Unlike Figure~\ref{fig:sing_duct_up} the boundary interface at  y = 21 \si{cm} appears smoother. 
  \begin{figure}[H]
\centering
\begin{minipage}{.45\textwidth}
  \centering
  \includegraphics[scale=0.5]{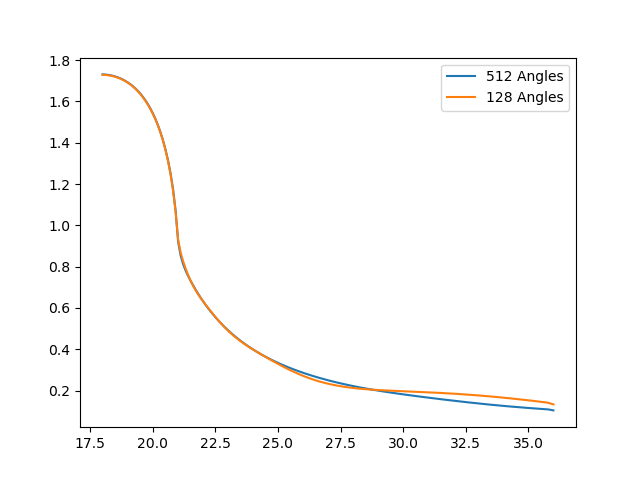}
  \subcaption{Scalar flux (\si{neutrons.cm^{-2}.s^{-1}}) vs y (\si{cm}) at x =14 \si{cm} generated using a neural network with Quadratic ConvFEM filters.}
  \label{fig:pg_angle_quad}
\end{minipage}%
\hfill
\begin{minipage}{.45\textwidth}
  \centering
  \includegraphics[scale=0.5]{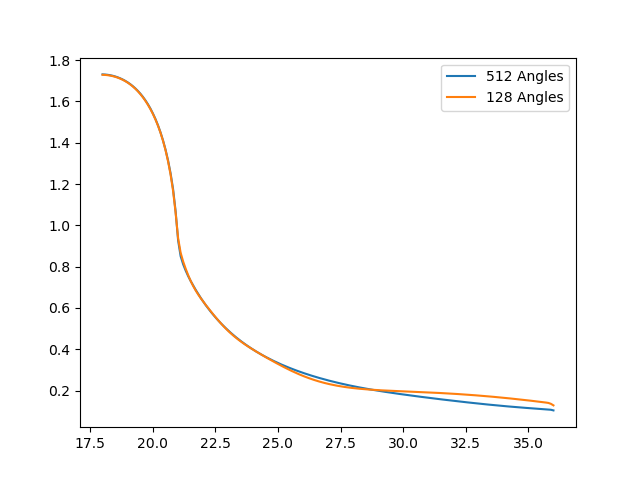}
  \subcaption{Scalar flux (\si{neutrons.cm^{-2}.s^{-1}}) vs y (\si{cm}) at x =14 \si{cm} generated using a neural network with Cubic ConvFEM filters.}
  \label{fig:pg_angle_quin}
\end{minipage}
\caption{Scalar Flux generated using a neural network with Quadratic and Cubic ConvFEM filters for $\Delta x = 0.1$ using 128 and 512 angular directions.}
\label{fig:pg_angle_comp}
\end{figure}
Figure~\ref{fig:pg_angle_comp} shows the Quadratic and Quintic ConvFEM filter solutions for $\Delta x = 0.1$ using 128 and 512 angles. It can be observed that the solution for 128 angles has a slight curve around the tail end, whereas the one with 512 angles does not have this. The solutions for $\Delta x = 0.05 \text{ and } 0.1$, therefore, use 512 for the subsequent comparisons. 

  \begin{figure}[H]
\centering
\begin{minipage}{.45\textwidth}
  \centering
  \includegraphics[scale=0.5]{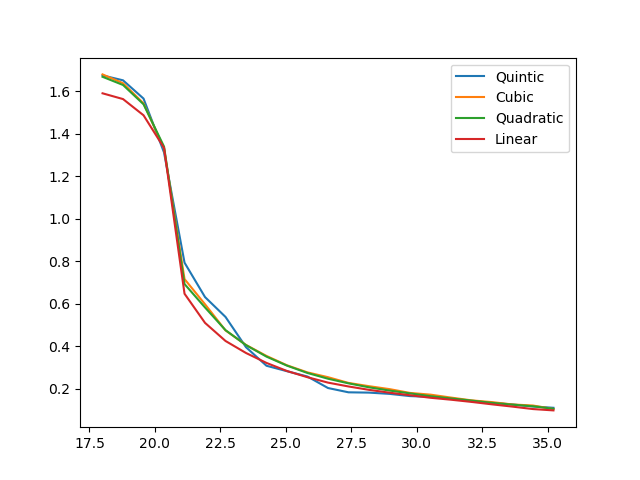}
  \subcaption{Scalar flux (\si{neutrons.cm^{-2}.s^{-1}}) vs y (\si{cm}) at x =14 \si{cm} for $\Delta x = 0.8$ using 128 angular directions.}
  \label{fig:pg_filter_08}
\end{minipage}%
\hfill
\begin{minipage}{.45\textwidth}
  \centering
  \includegraphics[scale=0.5]{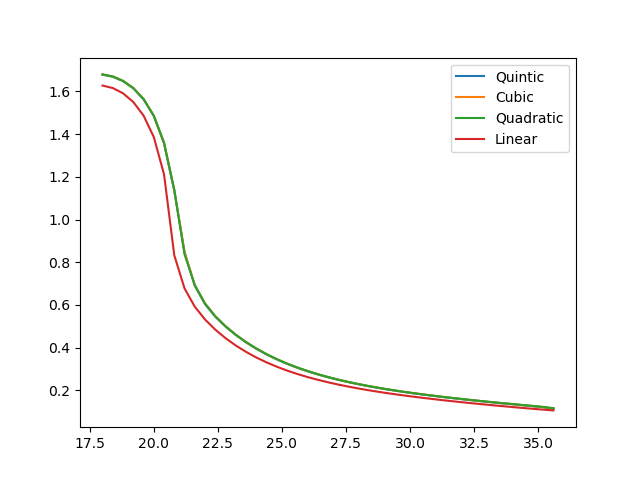}
  \subcaption{Scalar flux (\si{neutrons.cm^{-2}.s^{-1}}) vs y (\si{cm}) at x =14 \si{cm} for $\Delta x = 0.4$ using 128 angular directions.}
  \label{fig:pg_filter_04}
\end{minipage}
\caption{Scalar Flux generated using a neural network with Linear, Quadratic, Cubic and Quintic for $\Delta x = 0.8$ and $\Delta x = 0.4$.}
\label{fig:pg_filter_04_08}
\end{figure}
Figure~\ref{fig:pg_filter_04_08} shows the 1D flux profiles for all filters at $\Delta x = 0.8$ and $\Delta x = 0.4$. Figure~\ref{fig:pg_filter_08} is the 1D flux profiles for $\Delta x = 0.8$ and this shows an obvious difference between all filters, with the linear filter solution being much lower at the centre. When $\Delta x$ is decreased to $0.4$, the linear filter is the only one that shows a difference, with the quadratic, cubic and quintic all producing the same solution. 
  \begin{figure}[H]
\centering
\begin{minipage}{.45\textwidth}
  \centering
  \includegraphics[scale=0.5]{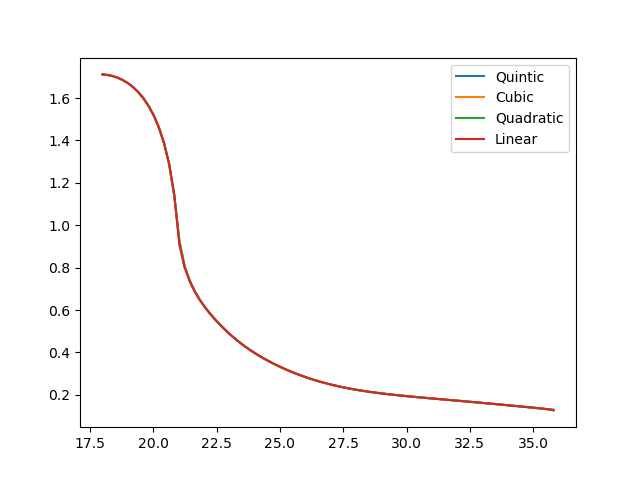}
  \subcaption{Scalar flux (\si{neutrons.cm^{-2}.s^{-1}}) vs y (\si{cm}) at x =14 \si{cm} for $\Delta x = 0.2$ using 128 angular directions.}
  \label{fig:pg_filter_02}
\end{minipage}%
\hfill
\begin{minipage}{.45\textwidth}
  \centering
  \includegraphics[scale=0.5]{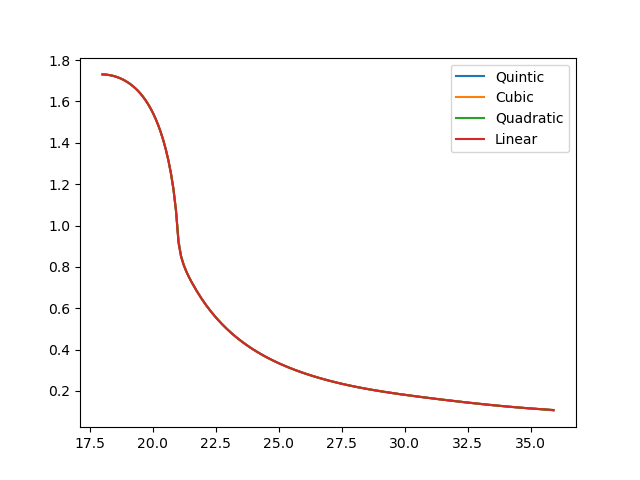}
  \subcaption{Scalar flux (\si{neutrons.cm^{-2}.s^{-1}}) vs y (\si{cm}) at x =14 \si{cm} for $\Delta x = 0.1$ using 512 angular directions.}
  \label{fig:pg_filter_01}
\end{minipage}
\caption{Scalar Flux generated using a neural network with Linear, Quadratic, Cubic and Quintic ConvFEM filters for $\Delta x = 0.2$ and $\Delta x = 0.1$.}
\label{fig:pg_filter_01_02}
\end{figure}
Figure~\ref{fig:pg_filter_01_02} shows the 1D flux profiles for all filters at $\Delta x = 0.2$ and $\Delta x = 0.1$. At these values of $\Delta x$ all four filters converge to the same solution. 
\begin{figure}[H]
    \centering
    \includegraphics[scale=0.7]{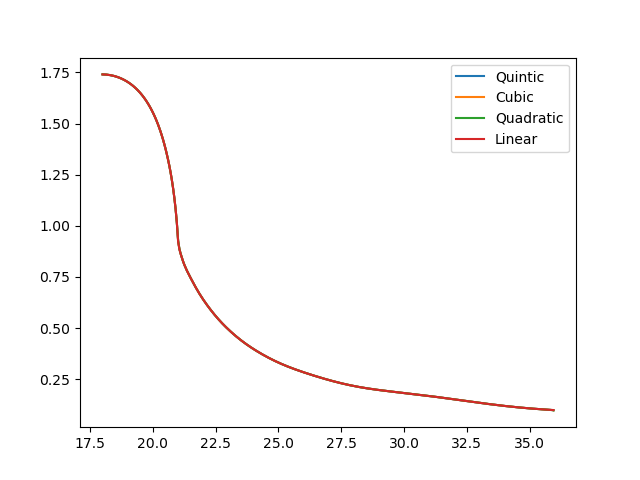}
    \caption{Scalar flux (\si{neutrons.cm^{-2}.s^{-1}}) vs y (\si{cm}) at x =14 \si{cm} for $\Delta x = 0.05$ using 512 angular directions.}
    \label{fig:pg_filter_005}
\end{figure}

Figure~\ref{fig:pg_filter_005} shows the 1D flux profiles for all filters at $\Delta x = 0.05$. Again, at these values of $\Delta x$ all four filters converge to the same solution. 

  \begin{figure}[H]
\centering
\begin{minipage}{.45\textwidth}
  \centering
  \includegraphics[scale=0.5]{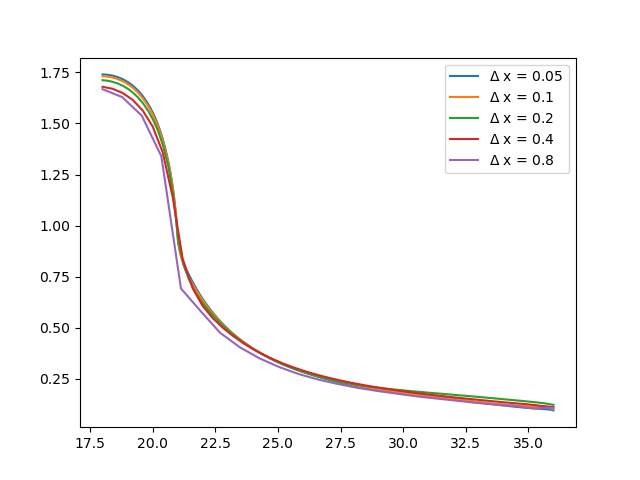}
  \subcaption{Scalar flux (\si{neutrons.cm^{-2}.s^{-1}}) vs y (\si{cm}) at x =14 \si{cm} generated using a neural network with Quadratic ConvFEM filters.}
  \label{fig:pg_filter_quad}
\end{minipage}%
\hfill
\begin{minipage}{.45\textwidth}
  \centering
  \includegraphics[scale=0.5]{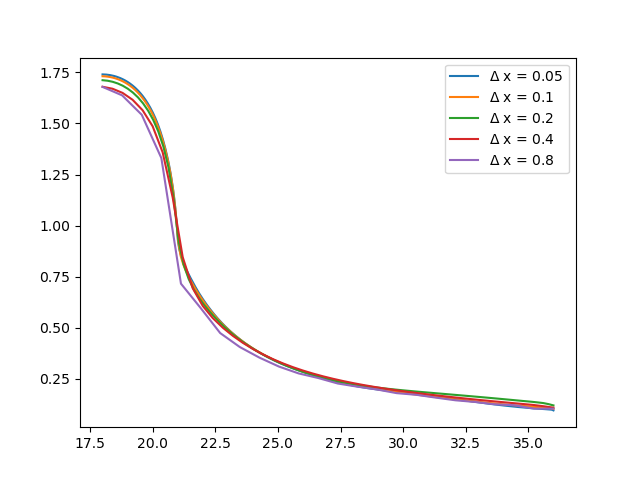}
  \subcaption{Scalar flux (\si{neutrons.cm^{-2}.s^{-1}}) vs y (\si{cm}) at x =14 \si{cm} generated using a neural network with Cubic ConvFEM filters.}
  \label{fig:pg_filter_cub}
\end{minipage}
\caption{Scalar Flux generated using a neural network with Quadratic and Cubic ConvFEM filters for $\Delta x = 0.05$ and $0.1$ using 512 angular directions and for $\Delta x = 0.2, 0.4$ and $0.8$ using 128 angular directions.} % using 512 angular directions.}
\label{fig:pg_filter_quad_cub}
\end{figure}

% \begin{figure}
\begin{figure}[H]
    \centering
    \includegraphics{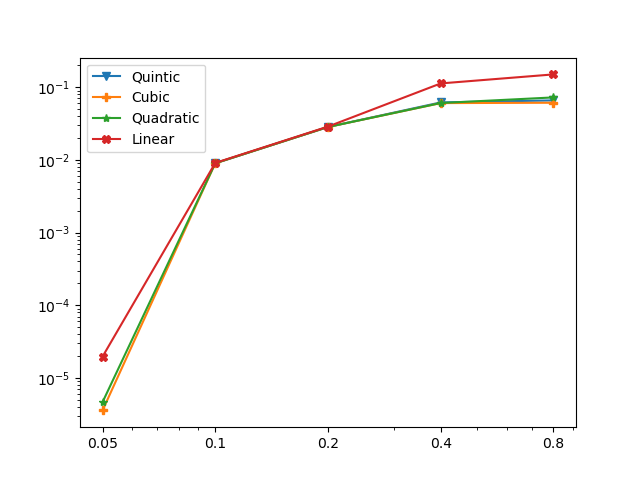}
    \caption{Error at  $x = 18$ \si{cm} and  $y = 14$ \si{cm} compared to the ConvFEM Quintic filter solution at $\Delta x = 0.05$ vs $\Delta x$}
    \label{fig:pd_comp_space}
\end{figure}

Figure~\ref{fig:pg_filter_quad} shows the Quadratic ConvFEM filter solutions for the four different spatial resolutions. As the spatial mesh is refined the peak flux in the centre tends to increase. The same pattern is seen when the cubic filters are used, shown in Figure~\ref{fig:pg_filter_cub}.
Figure~\ref{fig:pd_comp_space} shows the error at the domain centre: $x = 18 \si{cm}$ and  $y = 14 \si{cm}$ for each solution compared to the highest resolution solution, which is generated using the ConvFEM Quintic $9\times9$ filter with  $\Delta x = 0.05$. It can be observed that the error decreases as the order of the filter increases, with the exception of $\Delta x = 0.1$ and $0.2$.
%   \begin{figure}[H]
% \centering
% \begin{minipage}{.45\textwidth}
%   \centering
%   \includegraphics[scale=0.5]{figures/sing_duct_Pg_7/flux_2d.png}
%   \subcaption{Scalar flux (\si{neutrons.cm^{-2}.s^{-1}}) across the straight duct problem with the Petrov-Galerkin.}
%   \label{fig:fa_quad}
% \end{minipage}%
% \hfill
% \begin{minipage}{.45\textwidth}
%   \centering
%   \includegraphics[scale=0.5]{figures/sing_duct_Pg_7/flux_1d.png}
%   \subcaption{Scalar flux (\si{neutrons.cm^{-2}.s^{-1}}) vs y (\si{cm}) at x =70 \si{cm}.}
%   \label{fig:fa_keff}
% \end{minipage}
% \caption{Solution for the straight duct problem with the Petrov-Galerkin using Cubic FEM filters.}
% \label{fig:pg_cub}
% \end{figure}
\subsection{Fuel Assembly }\label{sec:fuel_assem_results}
The multi-group iteration network is now used to produce solutions for a 2D fuel assembly. This fuel assembly is based on the KAIST benchmark~\citep{kaist2000} and uses their cross-sections. 
\begin{figure}[H]
\centering
\scalebox{0.5}{\begin{tikzpicture}
\draw[step=1.0,black,thin] (0,0) grid (17,17);
\foreach \x in {0.5,3.5,6.5}
\foreach \y in {0.5,3.5,...,12.5}
\draw[black] (2+\y,5+\x) circle (7pt); 
\foreach \x in {0.5,3.5,6.5}
\foreach \y in {0.5,3.5,...,12.5}
\draw[black] (5+\x,2+\y) circle (7pt); 
\foreach \x in {3.5,13.5}
\foreach \y in {3.5,13.5}
\draw[black] (\x,\y) circle (7pt); 

\draw[step=1.0,black,thin] (18,11) grid (19,12);
\node[font=\huge,text width = 3cm] at (21,11.5) {Fuel Rod};
\draw[step=1.0,black,thin] (18,5) grid (19,6);
\draw[black] (18.5,5.5) circle (7pt); 
\node[font=\huge,text width = 5cm] at (22,5.5) {Guide-Tube or\\ Control Rod};
% \foreach \x in {0,...,14}
% \foreach \y in {0,...,14}
% \fill[green] (\x,\y) rectangle ++ (1,1); 

% \fill[blue] (1,1) rectangle ++ (1,1); 
% \fill[blue] (3,1) rectangle ++ (1,1); 
% \fill[blue] (1,3) rectangle ++ (1,1); 
% \fill[blue] (3,3) rectangle ++ (1,1); 

\end{tikzpicture}}
\caption{Geometry of UOX fuel assembly.}
\label{fig:fuel_assembly_mesh}
\end{figure}
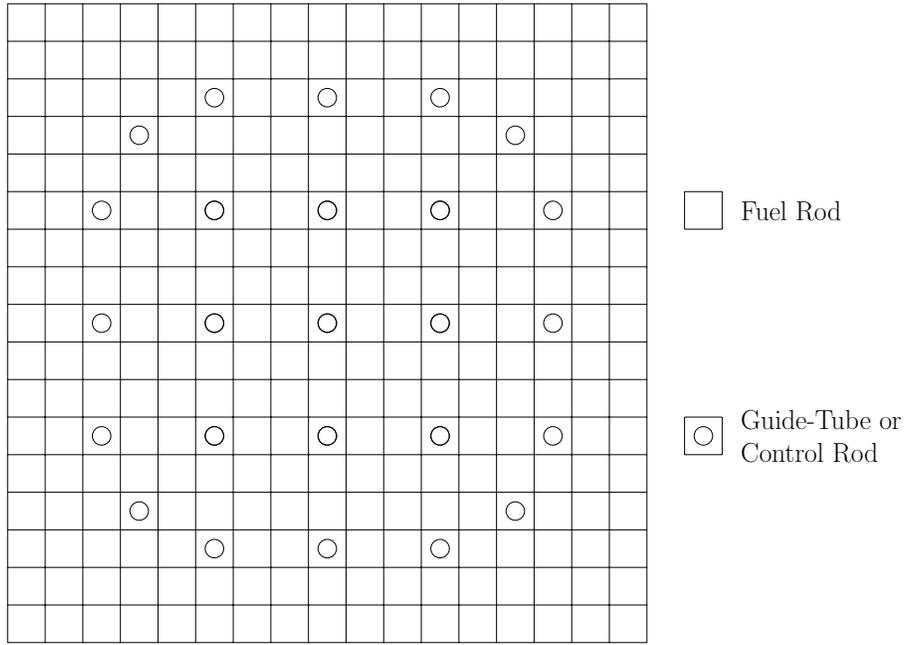

\begin{figure}[H]
\centering
\scalebox{0.5}{\begin{tikzpicture}
\draw[step=1,black,thin] (0,0) grid (8,8);

% \fill[lightgray] (\y,\x) rectangle ++ (0.5,0.5);

\foreach \x in {1,...,6}
\foreach \y in {3,4,...,4}
\fill[lightgray] (\x,\y) rectangle ++ (1,1);
\foreach \x in  {1,...,6}
\foreach \y in {3,4,...,4}
\fill[lightgray] (\y,\x) rectangle ++ (1,1);

\foreach \x in {2,...,5}
\foreach \y in {2,...,5}
\fill[lightgray] (\x,\y) rectangle ++ (1,1);
% \foreach \x in {2,3,...,7}
% \foreach \y in {14,15,...,17}
% \fill[lightgray] (\y,\x) rectangle ++ (1,1);

\draw[step=1.0,black,thin] (10,6) grid (11,7);
\node[font=\huge,text width = 3cm] at (13,6.5) {Moderator};
\draw[step=1.0,black,thin](10,2) grid (11,3);
\fill[lightgray] (10,2) rectangle ++ (1,1);
\node[font=\huge,text width = 5cm] at (14,2.5) {Guide, Control or Fuel };
% to make the diagram symmetrical so it looks better when centre-aligned
\node[font=\huge,text width = 5cm] at (-5,5.5) {\phantom{Guide-Tube or}};
% \foreach \x in {0,...,14}
% \foreach \y in {0,...,14}
% \fill[green] (\x,\y) rectangle ++ (1,1); 

% \fill[blue] (1,1) rectangle ++ (1,1); 
% \fill[blue] (3,1) rectangle ++ (1,1); 
% \fill[blue] (1,3) rectangle ++ (1,1); 
% \fill[blue] (3,3) rectangle ++ (1,1); 

\end{tikzpicture}}
\caption{Geometry of pins.}
\label{fig:fuel_pin_mesh}
\end{figure}
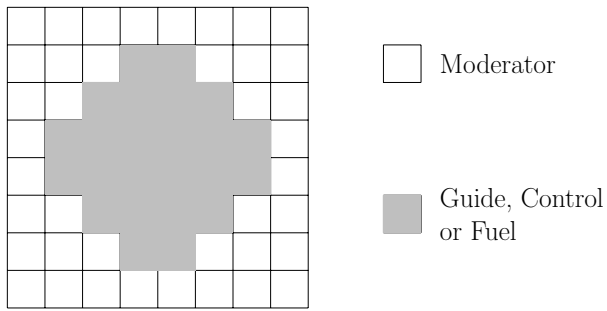

Figure~\ref{fig:fuel_assembly_mesh} contains the geometry for the UOX fuel assembly. This consists of a $17 \times 17$ lattice containing either UOX fuels rods, guide-tubes or control rods. Each square in the figure contains $8 \times 8$ cells/nodes, forming a total of $18,496$ cells along with $544$ cells to enforce the boundary conditions. The energy was discretised into seven groups, meaning the fuel assembly has $133,280$ degrees of freedom. Figure~\ref{fig:fuel_pin_mesh} contains the material parameters for the nodes in each pin.  Each side of the fuel assembly is length $21.42\text{cm}$ meaning each cell is $0.1575\text{cm} \times 0.1575\text{cm}$. Each side also has Vacuum boundary conditions applied to it. The material parameters for each pin are shown in figure~\ref{fig:fuel_pin_mesh}.
For solutions of the fuel assembly, two configurations are used to produce the solutions. Figure~\ref{fig:fuel_assembly_mesh} contains spaces that can either be guide-tubes or control rods. In the first configuration, all of these spaces are guide-tubes, representing a system where the control rods are fully withdrawn. In the second configuration, all of these spaces are control rods, representing a system where the control rods are fully inserted. This test case is multi-group and the power method~\cite{Golub1996} is the method chosen here to determine the dominant eigenvalue for this problem. The implementation of the power method used here is the same as~\cite{phillips2021} and the implementation of the multi-group network is the same as~\cite{Phillips2022diffusion}. For all solutions in this section, 3 Jacobi iterations, 100 multigrid iterations and 100 multi-group iterations are performed. 

  \begin{figure}[H]
\centering
\begin{minipage}{.45\textwidth}
  \centering
  \includegraphics[scale=0.6]{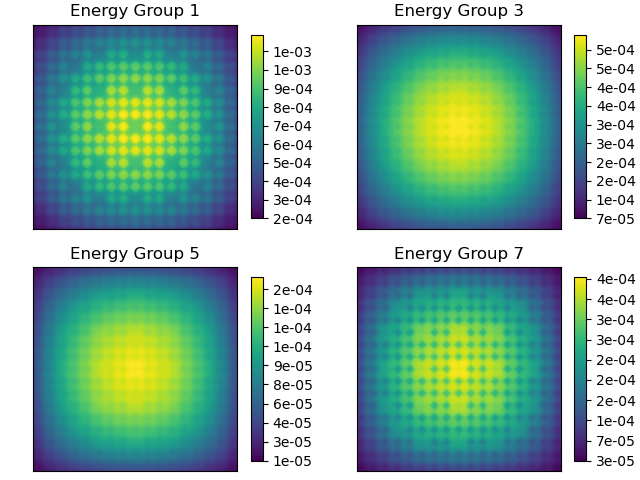}
  \subcaption{Scalar flux (\si{neutrons.cm^{-2}.s^{-1}}) across the fuel assembly for four energy groups for all control rods withdrawn.}
  \label{fig:fa_up_out_flux}
\end{minipage}%
\hfill
\begin{minipage}{.45\textwidth}
  \centering
  \includegraphics[scale=0.5]{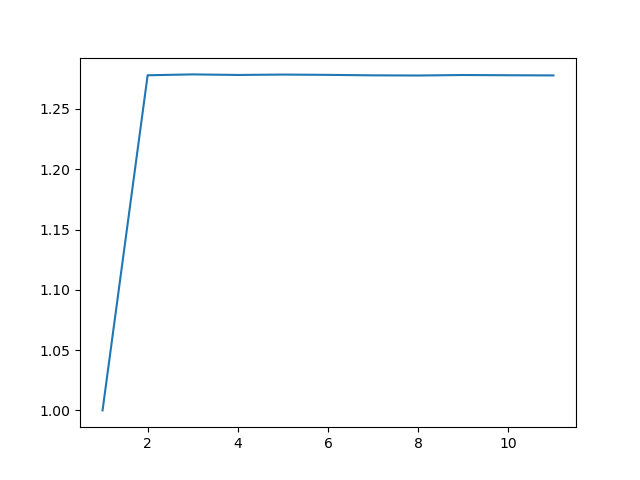}
  \subcaption{$k_{\text{eff}}$ convergence for a fuel assembly solution for all control rods withdrawn.}
  \label{fig:fa_up_out_keff}
\end{minipage}
\caption{Fuel assembly flux and $k_{\text{eff}}$ convergence for a fuel assembly solution with control rods fully withdrawn, generated using a neural network with the upwind method.}
\label{fig:fa_up_out_both}
\end{figure}

  \begin{figure}[H]
\centering
\begin{minipage}{.45\textwidth}
  \centering
  \includegraphics[scale=0.6]{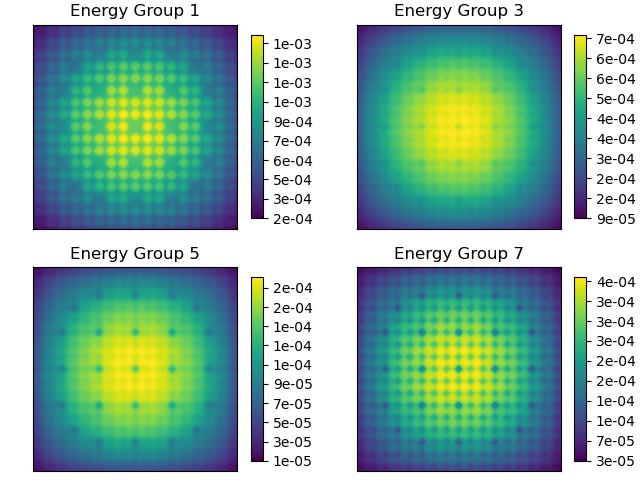}
  \subcaption{Scalar flux (\si{neutrons.cm^{-2}.s^{-1}}) across the fuel assembly for four energy groups for all control rods inserted.}
  \label{fig:fa_up_in_flux}
\end{minipage}%
\hfill
\begin{minipage}{.45\textwidth}
  \centering
  \includegraphics[scale=0.5]{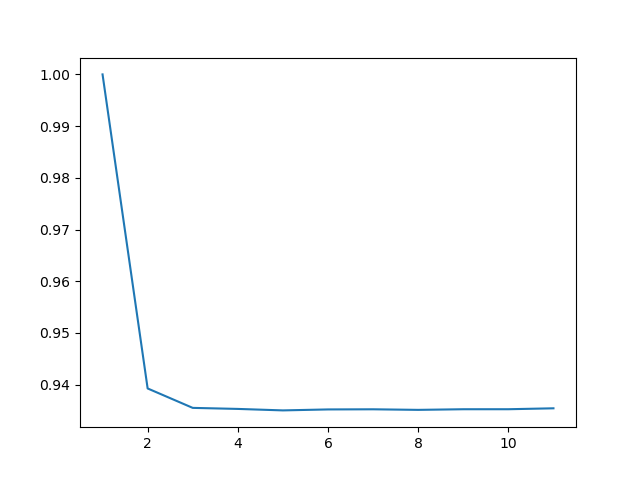}
  \subcaption{$k_{\text{eff}}$ convergence for a fuel assembly solution for all control rods inserted.}
  \label{fig:fa_up_in_keff}
\end{minipage}
\caption{Fuel assembly flux and $k_{\text{eff}}$ convergence for a fuel assembly solution with control rods fully inserted, generated using a neural network with the upwind method.}
\label{fig:fa_up_in_both}
\end{figure}

Figures~\ref{fig:fa_up_out_both} and~\ref{fig:fa_up_in_both} contain the scalar flux solution for a fuel assembly with control rods fully withdrawn and fully inserted respectively. Both solutions were generated using the upwind method.  $k_{\text{eff}}$ is notably lower when the control rods are fully inserted, as would be expected.

  \begin{figure}[H]
\centering
\begin{minipage}{.45\textwidth}
  \centering
  \includegraphics[scale=0.6]{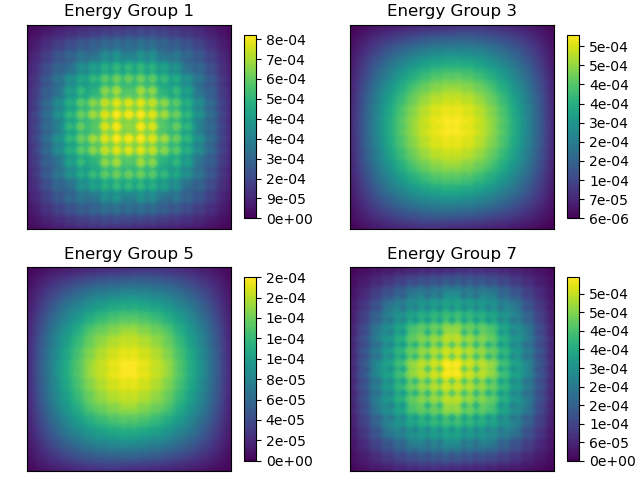}
  \subcaption{Scalar flux (\si{neutrons.cm^{-2}.s^{-1}}) across the fuel assembly for four energy groups for all control rods withdrawn.}
  \label{fig:fa_quad_out_quad}
\end{minipage}%
\hfill
\begin{minipage}{.45\textwidth}
  \centering
  \includegraphics[scale=0.5]{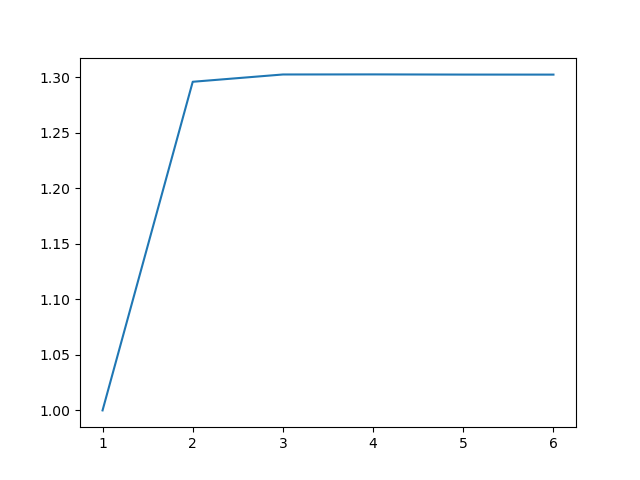}
  \subcaption{$k_{\text{eff}}$ vs power iteration for a fuel assembly solution for all control rods withdrawn.}
  \label{fig:fa_quad_out_quad}
\end{minipage}
\caption{Fuel assembly flux and $k_{\text{eff}}$ convergence for a fuel assembly solution with control rods fully withdrawn, generated using a neural network with Quadratic ConvFEM filters and the Petrov-Galerkin method.}
\label{fig:fa_quad_out_both}
\end{figure}

  \begin{figure}[H]
\centering
\begin{minipage}{.45\textwidth}
  \centering
  \includegraphics[scale=0.6]{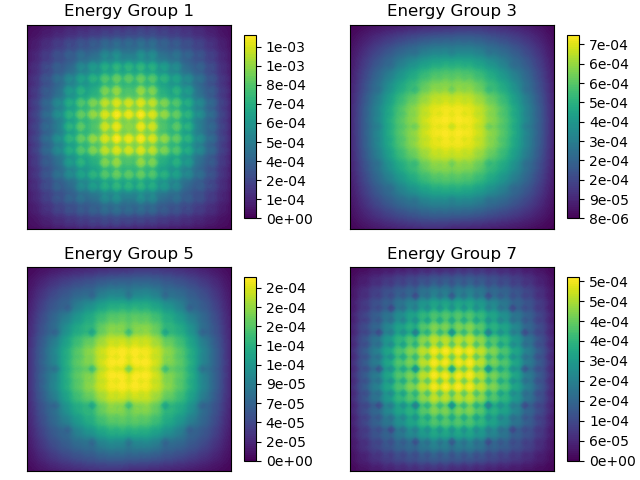}
  \subcaption{Scalar flux (\si{neutrons.cm^{-2}.s^{-1}}) across the fuel assembly for four energy groups for all control rods inserted.}
  \label{fig:fa_quad_in_quad}
\end{minipage}%
\hfill
\begin{minipage}{.45\textwidth}
  \centering
  \includegraphics[scale=0.5]{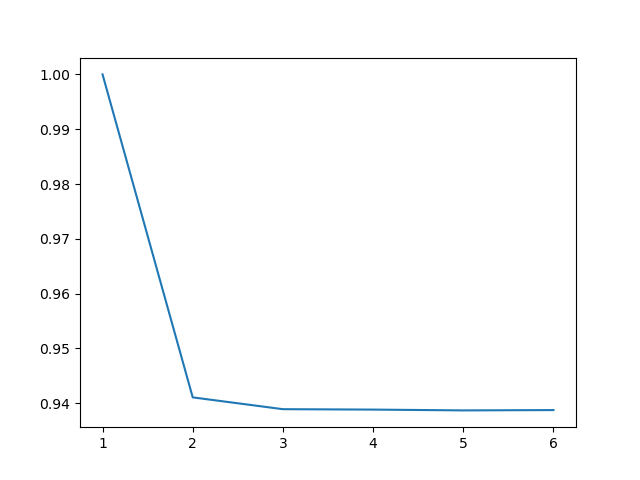}
  \subcaption{$k_{\text{eff}}$ vs power iteration for a fuel assembly solution for all control rods inserted.}
  \label{fig:fa_quad_in_keff}
\end{minipage}
\caption{Fuel assembly flux and $k_{\text{eff}}$ convergence for a fuel assembly solution with control rods fully inserted, generated using a neural network with Quadratic ConvFEM filters and the Petrov-Galerkin method.}
\label{fig:fa_quad_in_both}
\end{figure}
Figures~\ref{fig:fa_quad_out_both} and~\ref{fig:fa_quad_in_both} contain the scalar flux solution for a fuel assembly with control rods fully withdrawn and fully inserted respectively. Both solutions were generated using a neural network with Quadratic ConvFEM filters and the Petrov-Galerkin method. The solutions are similar to the upwind, although $k_{\text{eff}}$ is slightly higher for both solutions. 
The solution using the upwind method, Figures~\ref{fig:fa_up_out_both},\ref{fig:fa_up_in_both}, are similar to the Petrov-Galerkin method with quadratic ConvFEM filters, see Figures~\ref{fig:fa_quad_out_both},\ref{fig:fa_quad_in_both}. Although as expected the added dissipation associated with the upwind method results in a slight reduction in the maximum scalar flux in all the energy groups.

\section{Conclusions and Future work }\label{sec:conc}

In this paper, we introduce a new approach to solving the Boltzmann transport equation that is novel because it is  built using AI libraries.  
We develop the necessary methods in order to use AI libraries effectively, which includes a space-angle multigrid solution method.
% \textbf{(1) A space-angle multigrid solution method.}  
This can extract the level of parallelism necessary to run efficiently on GPUs or new AI computers.
% \textbf{(2) A new Convolutional Finite Element Method (ConvFEM)}.
A new Convolutional Finite Element Method (ConvFEM) is also developed. This enables one to have the same stencil at every node of the finite element mesh, unlike other high-order FEM approaches. 
This is important as it enables one to have simpler implementations using AI software, as the number of filters is greatly reduced. One would need a different filter for every FEM node stencil for conventional FEM approaches. The accuracy of the higher order (quadratic, cubic and quintic elements with $5\times5$, $7\times7$ and $9\times9$ stencils compared to the $3\times3$ stencils of linear FEM) methods was shown here although more work is necessary to fully demonstrate the accuracy of new ConvFEM. The linear FEM and ConvFEM are the same method and the only difference is between the higher order FEM and ConvFEM approaches. 
% \textbf{(3) A new non-linear Petrov-Galerkin method that introduces dissipation anisotropically.}  
Finally, a new non-linear Petrov-Galerkin method, that introduces dissipation anisotropically is developed.
This results in stable and accurate solutions. The solutions methods are not stable without the use of Petrov-Galerkin methods i.e. it can be difficult to achieve converged solutions of the Boltzmann transport discretised equations when purely central or Bubnov-Galerkin discretisations are used. 
Again more work is needed to fully explore the advantages of this approach.

There are several reasons why the use of AI libraries  is an attractive approach. For example, it enables one to use the highly optimised software within AI libraries, enabling one to run on different computer architectures and enabling one to tap into the vast quantity of community-based software that has been developed for AI and ML applications e.g. mixed arithmetic precision or model parallelism.  Also taking the lead from the massive neural networks that have been developed and the new AI computers, such as Cerebras 2 (with nearly a million cores) one can in principle use this approach for exascale computation.  In addition, since the overall method is essentially just a neural network experience with other neural network applications suggests that it can be combined or coupled with other neural networks that solve other equations e.g. fluid equations, to enable the solution of multi-physics problems. This is necessary to model a nuclear reactor with the various feedback mechanisms that feed into the nuclear criticality of the system e.g. temperature.  
We hope to explore all these promising areas in future work.  

Although the simulations shown here were run 
on GPUs and CPUs the next step would be to optimise the code and methods further so that large 3D problems can be run on GPUs or new AI computers and to test the performance of the methods on these computers. Extensions to unstructured meshes would also be a good future direction. 

\section*{CRediT authorship contribution statement}
%Conceptualization; 
%Methodology; 
%Software; 
%Validation; n/a
%Formal analysis; n/a
%Investigation; 
%Resources; 
%Data Curation; 
%Writing---Original Draft; 
%Writing---Review \& Editing; 
%Visualization; ...
%Supervision;
%Project administration; 
%Funding acquisition; 

\textbf{TRFP:} methodology, software, writing (original draft, review and editing).
\textbf{CEH:} methodology, writing (original draft, review and editing), supervision.
\textbf{BC:} software, writing (review and editing).
\textbf{AGB:} software, writing (original draft,  review and editing).
\textbf{CCP:} conceptualisation, methodology, software, writing (original draft, review and editing), supervision, funding acquisition.

\section*{Acknowledgements}
The authors would like to acknowledge the following EPSRC grants: RELIANT, Risk EvaLuatIon fAst iNtelligent Tool for COVID19 (EP/V036777/1); CO-TRACE, COvid-19 Transmission Risk Assessment Case Studies --- education Establishments (EP/W001411/1); INHALE,
Health assessment across biological length scales (EP/T003189/1); the PREMIERE programme grant (EP/T000414/1); MAGIC (EP/N010221/1); and MUFFINS (EP/P033180/1).

% \section*{References}

\bibliography{mybibfile}
\appendix
\section{The Convolution Finite Element Method (ConvFEM) Filters}
Here we assume a uniform grid with $\Delta x=\Delta y=$ constant and write the filters with this assumption. 
We also note that common to all filter orders is the $1\times1$ lumped mass filter $\bm{w_{ml}}=m_L=\Delta x \Delta y$. 
The filters below, as well as higher order filters, are listed in the GitHub repository~\cite{Phillips2022NNTS}. 
%~\url{https://github.com/trfphillips/Neural-Network-Transport-Solver}.
\subsection{Linear Filters}\label{app:linear_filters}
\begin{gather}
\bm{w_{x}}
 = \frac{1}{\Delta x}
 \begin{bmatrix*}[r]
 -0.167
 & 0 & 0.167 \\
-0.667 & 0 & 0.667 \\
 -0.167& 0 & 0.167
 \end{bmatrix*}
\end{gather}
\begin{gather}
\bm{w_{y}}
 = \frac{1}{\Delta y}
 \begin{bmatrix*}[r]
 0.167
 & 0.667 & 0.167 \\
0& 0 & 0 \\
 -0.167& -0.667 & -0.167
 \end{bmatrix*}
\end{gather}
\begin{gather}
\bm{w_{Diffxx}}
 = \frac{1}{(\Delta x)^2}
 \begin{bmatrix*}[r]
-0.167
 & 0.333 & -0.167 \\
-0.667& 1.333 & -0.667 \\
 -0.167& 0.333 & -0.167
 \end{bmatrix*}
\end{gather}
\begin{gather}
\bm{w_{Diffyy}}
 = \frac{1}{(\Delta y)^2}
 \begin{bmatrix*}[r]
-0.167
 & -0.667& -0.167 \\
0.333& 1.333 & 0.333 \\
 -0.167& -0.667 & -0.167
 \end{bmatrix*}
\end{gather}
\begin{gather}
\bm{w_{m}}
 = \frac{1}{(\Delta y)^2}
 \begin{bmatrix*}[r]
 \num{  2.78E-02 } & \num{  1.11E-01 } & \num{ 2.78E-02 } \\
\num{   1.11E-01 } & \num{  4.44E-01 } & \num{ 1.11E-01  } \\
\num{  2.78E-02 } & \num{  1.11E-01 } & \num{ 2.78E-02 }  
 \end{bmatrix*}
\end{gather}

\subsection{Quadratic Filters}\label{app:quad_filters}
\begin{gather}
\bm{w_{x}}
 = \frac{1}{\Delta x}
 \begin{bmatrix*}[r]
\num{-2.78E-03} & \num{	2.22E-02} & \num{	0.00E+00} & \num{	-2.22E-02} & \num{	2.78E-03} \\
\num{1.11E-02} & \num{	-8.89E-02} & \num{	0.00E+00} & \num{	8.89E-02} & \num{	-1.11E-02} \\
\num{6.67E-02} & \num{	-5.33E-01} & \num{	0.00E+00} & \num{	5.33E-01} & \num{	-6.67E-02} \\
\num{1.11E-02} & \num{	-8.89E-02} & \num{	0.00E+00} & \num{	8.89E-02} & \num{	-1.11E-02} \\
\num{-2.78E-03} & \num{	2.22E-02} & \num{	0.00E+00} & \num{	-2.22E-02} & \num{	2.78E-03}
 \end{bmatrix*}
\end{gather}
\begin{gather}
\bm{w_{y}}
 = \frac{1}{\Delta y}
 \begin{bmatrix*}[r]
\num{-2.78E-03} & \num{	1.11E-02} & \num{	6.67E-02} & \num{	1.11E-02} & \num{	-2.78E-03} \\
\num{2.22E-02} & \num{	-8.89E-02} & \num{	-5.33E-01} & \num{	-8.89E-02} & \num{	2.22E-02} \\
\num{0.00E+00} & \num{	0.00E+00} & \num{0.00E+00} & \num{	0.00E+00} & \num{	0.00E+00} \\
\num{-2.22E-02} & \num{	8.89E-02} & \num{	5.33E-01} & \num{	8.89E-02} & \num{	-2.22E-02} \\
\num{2.78E-03} & \num{	-1.11E-02} & \num{	-6.67E-02} & \num{	-1.11E-02} & \num{	2.78E-03}
 \end{bmatrix*}
\end{gather}

\begin{gather}
\bm{w_{Diffxx}}
 = \frac{1}{(\Delta x)^2}
 \begin{bmatrix*}[r]
\num{-2.78E-03} & \num{	4.44E-02} & \num{	-8.33E-02} & \num{	4.44E-02} & \num{	-2.78E-03} \\
\num{1.11E-02} & \num{	-1.78E-01} & \num{	3.33E-01} & \num{	-1.78E-01} & \num{	1.11E-02} \\
\num{6.67E-02} & \num{	-1.07E+00} & \num{	2.00E+00} & \num{	-1.07E+00} & \num{	6.67E-02} \\
\num{1.11E-02} & \num{	-1.78E-01} & \num{	3.33E-01} & \num{	-1.78E-01} & \num{	1.11E-02} \\
\num{-2.78E-03} & \num{	4.44E-02} & \num{	-8.33E-02} & \num{	4.44E-02} & \num{	-2.78E-03}
 \end{bmatrix*}
\end{gather}

\begin{gather}
\bm{w_{Diffyy}}
 = \frac{1}{(\Delta y)^2}
 \begin{bmatrix*}[r]
\num{-2.78E-03} & \num{	1.11E-02} & \num{	6.67E-02} & \num{	1.11E-02} & \num{	-2.78E-03} \\
\num{4.44E-02} & \num{	-1.78E-01} & \num{	-1.07E+00} & \num{	-1.78E-01} & \num{	4.44E-02} \\
\num{-8.33E-02} & \num{	3.33E-01} & \num{	2.00E+00} & \num{	3.33E-01} & \num{	-8.33E-02} \\
\num{4.44E-02} & \num{	-1.78E-01} & \num{	-1.07E+00} & \num{	-1.78E-01} & \num{	4.44E-02} \\
\num{-2.78E-03} & \num{	1.11E-02} & \num{	6.67E-02} & \num{	1.11E-02} & \num{	-2.78E-03}
 \end{bmatrix*}
\end{gather}
\begin{gather}
\bm{w_{m}}
 = \frac{1}{(\Delta y)^2}
 \begin{bmatrix*}[r]
   \num{1.11E-03} & \num{-4.44E-03} & \num{-2.66E-02} & \num{-4.44E-03} &  \num{1.11E-03} \\
  \num{-4.44E-03} &  \num{1.78E-02} & \num{1.07E-01}  & \num{1.78E-002} & \num{-4.44E-03} \\
  \num{-2.67E-02} & \num{1.07E-01}  & \num{6.40e-01} &  \num{1.07e-01}  & \num{-2.67E-02} \\
  \num{-4.44E-03}  & \num{1.78E-002} & \num{1.07E-01}   &     \num{1.78E-02} & \num{-4.44E-03} \\
   \num{1.11E-03} & \num{-4.44E-03} & \num{-2.67E-02}  &\num{-4.44E-03} &  \num{1.11E-03}
 \end{bmatrix*}
\end{gather}

\subsection{Cubic  Filters}\label{app:quintic_filters}
\begin{gather}
\bm{w_{x}}
 = \frac{1}{\Delta x}
 \begin{bmatrix*}[r]
\num{-3.30E-04} & \num{	2.26E-03} & \num{	-9.19E-03} & \num{	0.00E+00} & \num{	9.19E-03} & \num{	-2.26E-03} & \num{	3.30E-04}\\
\num{1.25E-03} & \num{	-8.57E-03} & \num{	3.48E-02} & \num{	0.00E+00} & \num{	-3.48E-02} & \num{	8.57E-03} & \num{	-1.25E-03}\\
\num{-2.03E-03} & \num{	1.39E-02} & \num{	-5.66E-02} & \num{	0.00E+00} & \num{	5.66E-02} & \num{	-1.39E-02} & \num{	2.03E-03}\\
\num{-2.69E-02} & \num{	1.85E-01} & \num{	-7.51E-01} & \num{	0.00E+00} & \num{	7.51E-01} & \num{	-1.85E-01} & \num{	2.69E-02}\\
\num{-2.03E-03} & \num{	1.39E-02} & \num{	-5.66E-02} & \num{	0.00E+00} & \num{	5.66E-02} & \num{	-1.39E-02} & \num{	2.03E-03}\\
\num{1.25E-03} & \num{	-8.57E-03} & \num{	3.48E-02} & \num{	0.00E+00} & \num{	-3.48E-02} & \num{	8.57E-03} & \num{	-1.25E-03}\\
\num{-3.30E-04} & \num{	2.26E-03} & \num{	-9.19E-03} & \num{	0.00E+00} & \num{	9.19E-03} & \num{	-2.26E-03} & \num{	3.30E-04}
 \end{bmatrix*}
\end{gather}
and similarly for $\bm{w_y}$ etc.

\end{document}